\documentclass[journal]{IEEEtran}

\ifCLASSINFOpdf
\else
\usepackage[dvips]{graphicx}
\fi

\usepackage{booktabs}

\usepackage{amsmath,epsfig,amssymb,amsthm,url,amsfonts}
\usepackage{algorithm}
\usepackage{algpseudocode}
\usepackage{multirow}
\usepackage{mdframed}
\usepackage{url}
\usepackage{enumitem}
\usepackage[makeroom]{cancel}
\usepackage{dblfloatfix}
\usepackage{array}
\usepackage{epstopdf}
\usepackage{float}
\usepackage{subfig}
\usepackage{breqn}
\usepackage{cite}
\usepackage{xcolor}
\usepackage{tabularx}
\usepackage[printonlyused,withpage]{acronym}
\usepackage{balance}
\usepackage{graphbox} 
\newtheorem{lemma}{Lemma}
\setlist[itemize]{label=$\triangleright$}
\newtheoremstyle{break}
{}
{}
{\itshape}
{}
{\bfseries}
{.}
{\newline}
{}
\theoremstyle{break}

\theoremstyle{definition}

\DeclareMathOperator{\trace}{tr}

\DeclareMathOperator{\vect}{Vect}

\newcommand{\E}{\mathbb{E}}
\newcommand{\bs}{\boldsymbol}

\def\mbb#1{\mathbb{#1}}
\def\bs#1{\boldsymbol{#1}}

\makeatletter
\def\thmhead@plain#1#2#3{%
	\thmname{#1}\thmnumber{\@ifnotempty{#1}{ }\@upn{#2}}%
	\thmnote{ {\the\thm@notefont#3}}}
\let\thmhead\thmhead@plain
\makeatother

\graphicspath{{./figs/}}

\newcommand{\argmin}{\operatornamewithlimits{argmin}}

\newcommand{\lk}{ \left\{ }
\newcommand{\rk}{ \right\} }

\newcommand{\tr}{ {\mbox{{trace}}} }

\newcommand{\Rbb}{{\mathbb{R}}}

\newcommand{\Hb}{{\bf H}}

\newcommand{\bb}{{\bf b}}

\newcommand{\sbb}{{\bf s}}

\newcommand{\xb}{{\bf x}}

\newcommand{\zb}{{\bf z}}

\newcommand{\Wb}{{\bf W}}

\newcommand{\cc}{{\cal c}}

\newcommand{\Sigb}{{\mbox{\boldmath $\Sigma$}}}

\newcommand{\Ib}{{\bf I}}

\newcommand{\mub}{{\mbox{\boldmath $\mu$}}}

\newcommand{\alphab}{{\mbox{\boldmath $\alpha$}}}

\newcommand{\vb}{{\mathbf v}}
\newcommand{\Vb}{{\mathbf V}}

\newcommand{\Nc}{{\cal N}}

\newsavebox\mybox

\acrodef{SE}{speech enhancement}
\acrodef{STFT}{short-time Fourier transform}
\acrodef{STOI}{short-time objective intelligibility}
\acrodef{PSD}{power spectral density}
\acrodef{NMF}{nonnegative matrix factorization}
\acrodef{AV}{audio-visual}
\acrodef{DNN}{deep neural network}
\acrodefplural{DNNs}{deep neural networks}
\acrodef{VAE}{variational auto-encoder}
\acrodefplural{VAEs}{variational auto-encoders}
\acrodef{CVAE}{conditional variational auto-encoder}
\acrodefplural{CVAEs}{conditional variational auto-encoders}
\acrodef{A-VAE}{audio VAE}
\acrodef{V-VAE}{visual VAE}
\acrodef{AV-CVAE}{audio-visual CVAE}
\acrodef{ROI}{region of interest}
\acrodef{MCMC}{Markov Chain Monte Carlo}
\acrodef{EM}{expectation-maximization}
\acrodef{MCEM}{Monte Carlo expectation-maximization}
\acrodef{TF}{time frequency}
\acrodef{ELBO}{evidence lower bound}
\acrodef{ROI}{region of interest}
\acrodef{LR}{Living Room}
\acrodef{SDR}{signal-to-distortion ratio}
\acrodef{PESQ}{perceptual evaluation of speech quality}
\acrodef{ASE}{audio speech enhancement}
\acrodef{VSE}{visual speech enhancement}
\acrodef{AVSE}{audio-visual speech enhancement}
\acrodef{SNR}{signal-to-noise ratio}
\acrodefplural{SNRs}{signal-to-noise ratios}
\acrodef{LSTM}{long short-term memory}
\acrodef{DNNs}{deep neural networks}

\begin{document}
	
	\title{Mixture of Inference Networks for VAE-based Audio-visual Speech Enhancement}
	
	\author{Mostafa Sadeghi and Xavier Alameda-Pineda, \IEEEmembership{Senior Member, IEEE}

		\thanks{Mostafa Sadeghi is with the Multispeech team at Inria Nancy - Grand Est, France. Xavier Alameda-Pineda is with the Perception team at Inria Grenoble Rh\^{o}ne-Alpes and Universit\'e Grenoble Alpes, France.}
				\thanks{Xavier Alameda-Pineda acknowledges the ANR ML3RI (ANR-19-CE33-0008-01) and the ANR-3IA MIAI (ANR-19-P3IA-0003).}
		}
	
	\maketitle
	
	\begin{abstract}
		In this paper, we are interested in unsupervised (unknown noise) audio-visual speech enhancement based on variational autoencoders (VAEs), where the probability distribution of clean speech spectra is simulated using an encoder-decoder architecture. The trained generative model (decoder) is then combined with a noise model at test time to estimate the clean speech. In the speech enhancement phase (test time), the initialization of the latent variables, which describe the generative process of clean speech via decoder, is crucial, as the overall inference problem is non-convex. This is usually done by using the output of the trained encoder where the noisy audio and clean visual data are given as input. Current audio-visual VAE models do not provide an effective initialization because the two modalities are tightly coupled (concatenated) in the associated architectures. To overcome this issue, inspired by mixture models, we introduce the mixture of inference networks variational autoencoder (MIN-VAE). Two encoder networks input, respectively, audio and visual data, and the posterior of the latent variables is modeled as a mixture of two Gaussian distributions output from each encoder network. The mixture variable is also latent, and therefore the inference of learning the optimal balance between the audio and visual inference networks is unsupervised as well. By training a shared decoder, the overall network learns to adaptively fuse the two modalities. Moreover, at test time, the visual encoder, which takes (clean) visual data, is used for initialization. A variational inference approach is derived to train the proposed generative model. Thanks to the novel inference procedure and the robust initialization, the proposed MIN-VAE exhibits superior performance on speech enhancement than using the standard audio-only as well as audio-visual counterparts.
	\end{abstract}
	
	\IEEEpeerreviewmaketitle
	
	\section{Introduction}
	\IEEEPARstart{S}{peech} enhancement, or removing background noise from noisy speech~\cite{VincVG18,loizou2007speech}, is a classic yet very important problem in signal processing and machine learning. Traditional solutions to this problem are based on spectral subtraction~\cite{boll1979suppression} and Wiener filtering~\cite{lim1979enhancement}, targeting noise and/or speech \ac{PSD} estimation in the \ac{STFT} domain. The recent impressive performance of \ac{DNNs} in computer vision and machine learning has paved the way to revisit the speech enhancement problem. \ac{DNNs} have been widely utilized in this regard, where a neural network is trained to map a noisy speech spectrogram to its clean version, or to a \ac{TF} mask~\cite{wang2018supervised,xu2015regression,li:hal-02264247}. This is usually done in a supervised way, using a huge dataset of noise and clean speech signals for training. As such, the performance of a supervised speech enhancement technique often degrades when dealing with an unknown type of noise.
	
	{Unsupervised techniques provide another procedure for speech enhancement that does not use noise signals for training. A popular unsupervised method is based on \ac{NMF} ~\cite{wilson2008speech, mohammadiha2013supervised, SediBRJ17} for modeling the \ac{PSD} of speech signals~\cite{ISNMF}, which decomposes PSD as a product of two non-negative low-rank matrices (a dictionary of basis spectra and the corresponding activations). An NMF-based speech enhancement method consists of first learning a set of basis spectra for clean speech spectrograms at training phase, prior to speech enhancement \cite{SmarRS07_nmf, MysoS11, mohammadiha2013supervised}. Then, by decomposing the noisy spectrogram as the sum of clean speech and noise spectrograms, the corresponding clean speech activations as well as the NMF parameters of noise are estimated. While being computationally efficient, this modeling and enhancement framework cannot properly explain complicated structure of speech spectrogram due to the limited representational power dictated by the two low-rank matrices. A deep autoencoder (DAE) has been employed in \cite{SunZZ16} to model clean speech and noise spectrograms. A DAE is pre-trained for clean speech spectrograms, while an extra DAE for noise spectrogram is trained at the enhancement stage using the noisy spectrogram. The corresponding inference problem is under-determined, and the authors proposed to constrain the unknown speech using a pre-trained NMF model. As such, this DAE-based method might encounter the same shortcomings as those of the NMF-based speech enhancement \cite{bando2018statistical}.}
	
	Deep latent variable models offer a more sophisticated and efficient modeling framework than NMF and DAE, gaining much interest over the past few years~\cite{bando2018statistical,Leglaive_MLSP18,SekiguchiAPSIPA2018,PariDV19,Leglaive_ICASSP2019b,KameLIM19,nguyen2020deep}. The first and main step is to train a generative model for clean speech spectrogram using a \ac{VAE}~\cite{RezeMW14,KingW14}. \ac{VAE} provides an efficient way to estimate the parameters of a non-linear generative model, also called the decoder. This is done by approximating the intractable posterior distribution of the latent variables using a Gaussian distribution parametrized by a neural network, called the inference (encoder) network. The encoder and decoder are jointly trained to maximize a variational lower bound on the marginal data log-likelihood. At test time, the trained generative model is combined with a noise model, e.g.\ \ac{NMF}. The unknown noise parameters and clean speech are then estimated from the observed noisy speech. {Being independent of the noise type at training, these methods show better generalization than the supervised approaches~\cite{bando2018statistical,Leglaive_MLSP18}. }
	
	Motivated by the fact that the visual information, when associated with audio information, often helps improve the performance in various tasks~\cite{cech2013active,ban2019variational,ban2017tracking}, and in particular the quality of speech enhancement~\cite{girin2001audio,AfouCZ18,GabbSP18}, an audio-visual latent variable generative model has recently been proposed in~\cite{sadeghiLAGH19}. Within this model, the visual features corresponding to the lips region of the speaker are also fed to the encoder and decoder networks of the VAE. {The effectiveness and superior performance of the audio-visual VAE (AV-VAE) compared to the audio-only VAE (A-VAE), as well as the supervised deep learning based method of \cite{GabbSP18} has been experimentally verified in~\cite{sadeghiLAGH19}.} To deal with noisy visual data at test time, e.g. non-frontal or occluded lips images, a robust method has been proposed in~\cite{SadeA19a}, which was later improved and extended in \cite{SadeA21Switching}.  In the proposed approach, a mixture of trained A-VAE and AV-VAE is used as the clean speech model during speech enhancement. Because of that, the deteriorating effects associated with missing/noisy visual information are avoided as the algorithm switches from AV-VAE to A-VAE in these cases~\cite{SadeA19a}. Besides AV-VAE, a video-only VAE (V-VAE) has also been introduced in~\cite{sadeghiLAGH19}, where the posterior parameters of the latent variables, that is, the encoder parameters, are trained using only visual information. As such, the latent variables governing the generative process of clean speech spectrogram are inferred from visual data only. V-VAE has been shown to yield much better speech enhancement performance than A-VAE when the (acoustic) noise level is high~\cite{sadeghiLAGH19}. 
	
	In the speech enhancement phase, because of the non-linear generative model, the posterior of the latent variables does not admit a closed-form expression. Two approaches are often used to get around this problem. The first solution is based on the Markov Chain Monte
	Carlo (MCMC) method~\cite{bishop06}, in which a sampling technique, e.g. the Metropolis-Hastings algorithm~\cite{bishop06}, is used to sample from the posterior~\cite{bando2018statistical,Leglaive_MLSP18}. The obtained samples are then used to approximate the expectations using a Monte-Carlo average. The second approach makes use of optimization techniques to find the maximum a posteriori estimation of the latent variables~\cite{KameLIM19}. In either case, the initialization plays an important role, as the associated problems are highly non-convex. In practice, the trained encoder is used to initialize the latent variables by giving the noisy speech spectrogram as the input and taking the posterior mean at the output. This can partly explain why V-VAE performs better than A-AVE at high noise levels. In fact, the latent variable initialization in V-VAE is based on visual features, whereas in A-VAE, it is based on the noisy speech. As a result, V-VAE provides a better initialization, because it uses noise-free data (visual features)~\cite{sadeghiLAGH19}.
	
	The original contribution of this paper is to optimally exploit the complementarity of A-VAE and V-VAE, without systematic recourse to simultaneously using audio and visual features, i.e. via simple concatenation (tight fusion) as done in AV-VAE. Indeed, we aim to bridge the performance gap between A-VAE and V-VAE by designing a mixture of audio and visual inference (encoder) networks, called mixture of inference networks VAE (MIN-VAE). The inputs to audio and visual encoders are speech spectrogram frames and the corresponding visual features, respectively, thus training MIN-VAE to select the best combination of the the audio and visual information. A variational inference approach is proposed to train the mixture of the two encoders jointly with a shared decoder (generative) network. This way, the decoder reconstructs the input audio data using the optimal combination of the audio and visual latent samples. At test time, the latent variables are initialized using the visual encoder, thus providing a robust initialization. Our experiments show that MIN-VAE yields much better performance than previous methods, i.e. A-VAE, V-VAE, and AV-VAE.
	
	It should be noted that there are some fundamental differences between our proposed MIN-VAE and the mixture model introduced in \cite{SadeA19a}. While the purpose of our work is to combine an A-VAE with a V-VAE to take advantage of the both in terms of robust initialization and an improved generative model, the work in \cite{SadeA19a} addresses robust audio-visual speech enhancement. We achieve our goal by proposing a VAE architecture with a single decoder but a mixture of audio- and visual-based encoders. A new inference method is also derived to train the proposed VAE. In \cite{SadeA19a}, the robustness is achieved by considering a mixture of an A-VAE's decoder and an AV-VAE's decoder at test phase. Both the decoders have been trained separately (using standard A-VAE and AV-VAE), and no particular VAE architecture is trained. In contrast to our present work, the architecture proposed in \cite{SadeA19a} does not provide robustness to latent initialization as it uses a VAE architecture where the audio and visual modalities are tightly fused.
	
	The rest of the paper is organized as follows. In Section~\ref{sec:sse}, we review clean speech modeling using already proposed VAE architectures. Next, Section~\ref{sec:prop} introduces our proposed MIN-VAE modeling and the associated speech enhancement strategy. Experimental results are then presented in Section~\ref{sec:exp}.
	\begin{figure*}[t!]
		\centering
		\rotatebox[origin=c]{90}{A-VAE}\hspace{8mm}\includegraphics[align=c,width=0.70\textwidth]{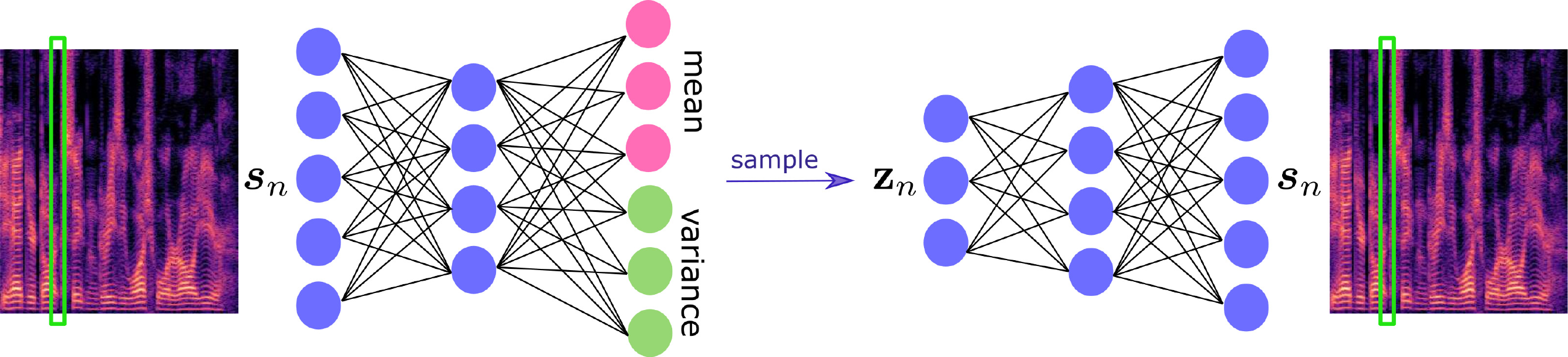}\vspace{2mm}
		\hrule
		\vspace{2mm}
		\rotatebox[origin=c]{90}{V-VAE}\hspace{8mm}\includegraphics[align=c,width=0.75\textwidth]{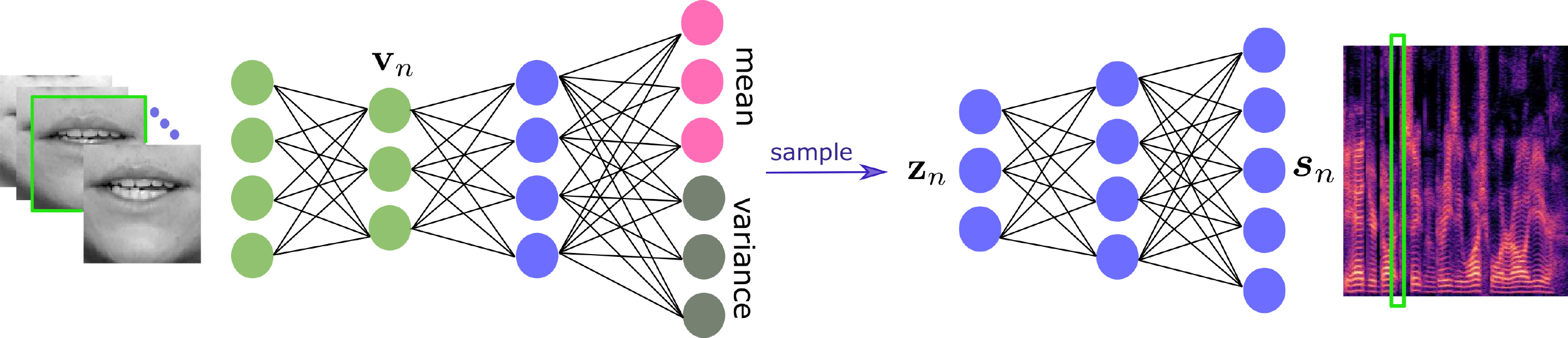}\vspace{2mm}
		\hrule
		\vspace{2mm}
		\rotatebox[origin=c]{90}{AV-VAE}\hspace{8mm}\includegraphics[align=c,width=0.85\textwidth]{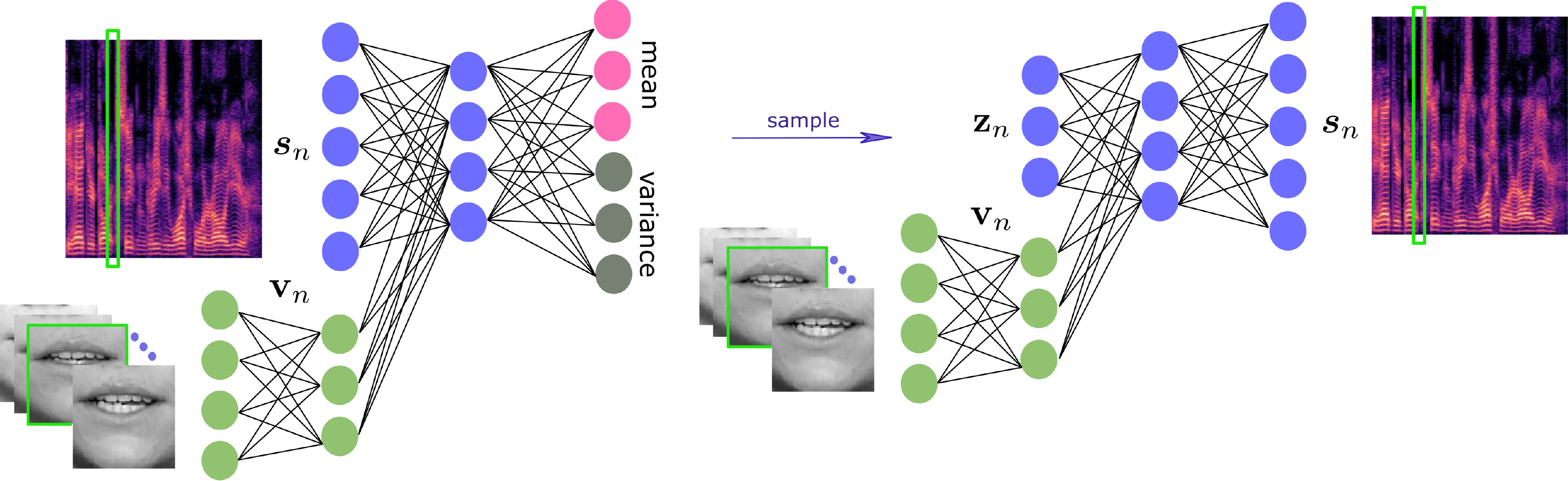}
		\caption{Architectures for (top) the Audio-only VAE (A-VAE) proposed in~\cite{Leglaive_MLSP18}, (middle) the Video-only VAE (V-VAE) proposed in~\cite{sadeghiLAGH19} and (bottom) the Audio-Visual VAE (AV-VAE) proposed in~\cite{sadeghiLAGH19}.}
		\label{fig:vae_architectures}
	\end{figure*}
	\section{VAE-based Speech Modeling}\label{sec:sse}
	In this section, audio-only, visual-only and audio-visual clean speech modeling based on VAE is reviewed. Roughly speaking, this consists in defining a latent variable generative model for each time frame of clean speech spectrogram. A parametric Gaussian distribution is  used to define the conditional distribution of a spectrogram time frame given its associated latent variable (and, depending on the choice, the visual feature vector). The parameters of the distribution are modeled by \ac{DNNs}. Assuming a standard Gaussian prior distribution for the latent variables, the model (DNN) parameters are then learned from a collection of clean training data using variational inference. To do so, the posterior distribution of the latent variables is approximated by a Gaussian distribution parametrized by a DNN, called the encoder network. In what follows, three VAE-based modeling frameworks are reviewed.
	\subsection{Audio-only VAE}
	Let $ \sbb_n\in \mathbb{C}^F $ denote the vector of speech \ac{STFT} coefficients at time frame $ n $, for $n \in \{0,...,N-1\}$, which is assumed to be generated according to the following latent variable model~\cite{bando2018statistical,Leglaive_MLSP18}:
	\begin{align}
		\label{decoder_VAE}
		\sbb_n | \mathbf{z}_n &\sim \mathcal{N}_c\Big(\boldsymbol{0}, \text{diag}\Big(\bs{\sigma}_{s}(\mathbf{z}_n)\Big)\Big), \\
		\label{prior_VAE}
		\mathbf{z}_n &\sim \mathcal{N}(\mathbf{0}, \mathbf{I}), 
	\end{align}
	where $\mathbf{z}_n \in \mathbb{R}^L$, with $L \ll F$, is a latent random variable, 
	$\mathcal{N}_c(\mathbf{0}, \Sigb)$ denotes a
	zero-mean complex proper Gaussian distribution with covariance matrix $\Sigb$, and $\mathcal{N}(\mathbf{0}, \mathbf{I})$ stands for a zero-mean Gaussian distribution with an identity covariance matrix. Moreover, $ \bs{\sigma}_{s}(.): \mbb{R}^L \mapsto 
	\mbb{R}_+^F $ is modeled with a neural network parameterized by $ \bs{\theta} $, which is called the \textit{decoder}.
	
	In order to estimate the set of parameters, $ \bs{\theta} $, the \ac{VAE} formalism proposes to approximate the intractable posterior distribution $ p(\mathbf{z}_n |\mathbf{s}_n) $ by a variational distribution parametrized by a neural network, called the \textit{inference (encoder) network}~\cite{KingW14}. This variational distribution writes:
	\begin{equation}
		\label{eq:varapp}
		q(\mathbf{z}_n |\mathbf{s}_n;\bs{\phi}_a)= \mathcal{N}\Big(\bs{\mu}_z^a(\sbb_n), \text{diag}\Big(\bs{\sigma}_{z}^a(\mathbf{s}_n)\Big)\Big),
	\end{equation}
	where, $ \bs{\mu}_z^a(.): \mbb{R}_{+}^F \mapsto 
	\mbb{R}^L $ and $ \bs{\sigma}_{z}^a(.): \mbb{R}_{+}^F \mapsto 
	\mbb{R}_+^L $ are neural networks, with parameters denoted $ \bs{\phi}_a $, taking $ \tilde{\mathbf{s}}_n \triangleq (|s_{0n}|^2 \dots |s_{F-1\: n}|^2)^{\top}$ 
	as input. Given a sequence of STFT speech time frames $ \sbb=\lk \sbb_n\rk_{n=0}^{N_{tr}-1} $ as training data, with $ \zb=\lk \zb_n\rk_{n=0}^{N_{tr}-1} $ being the associated latent variables, the parameters $ \lk\bs{\theta},\bs{\phi}_a\rk $ are then estimated by maximizing a lower bound on the data log-likelihood $ \log p(\sbb;\bs{\theta}) $. Note that,
	\begin{align}
		\log p(\sbb;\bs{\theta}) &=\log \int p(\sbb|\zb;\bs{\theta})p(\zb)d\zb\nonumber\\
		&\ge \E_{q(\zb|\mathbf{s};\bs{\phi}_a)}\Big[ \log \frac{p(\sbb|\zb;\bs{\theta})p(\zb)}{q(\zb|\mathbf{s};\bs{\phi}_a)} \Big] \triangleq \mathcal{L}\left(\bs{\theta}, \bs{\phi}_a \right)
	\end{align}
	where, the Jensen's inequality has been used, as it is classically done, see~\cite{KingW14}. The function $ \mathcal{L}\left(\bs{\theta}, \bs{\phi}_a \right) $ is called the evidence lower bound 
	(ELBO)~\cite{KingW14}, because it provides a lower bound on $ \log p(\sbb;\bs{\theta}) $. The ELBO can be decomposed as:
	\begin{multline}
		\label{eq:elbo_vae}
		\mathcal{L}\left(\bs{\theta}, \bs{\phi}_a \right) = \E_{q(\zb|\mathbf{s};\bs{\phi}_a)}\Big[ \ln p\left(\mathbf{s} | \mathbf{z} ; \bs{\theta} \right) \Big]  -\\{{\cal D}_{KL}}\Big(q\left(\mathbf{z}| \mathbf{s} ; \bs{\phi}_a\right) \parallel p(\mathbf{z})\Big),
	\end{multline}
	where, $ {{\cal D}_{KL}}(q\parallel p) $ denotes the Kullback-Leibler (KL)
	divergence between $ q $ and $ p $. The first term in the right-hand side of \eqref{eq:elbo_vae} evaluates the reconstruction quality of the decoder, and the second one is a regularization term encouraging the variational posterior to remain close to the prior. As the expectation in \eqref{eq:elbo_vae} is computationally intractable, it is usually approximated by a single sample drawn from $ q(\zb|\mathbf{s};\bs{\phi}_a) $~\cite{KingW14}. Employing a so-called re-parameterization trick, the set of parameters $ \lk\bs{\theta},\bs{\phi}_a\rk $ is estimated by a stochastic gradient ascent algorithm~\cite{KingW14}. Since all the parameters are inferred using only audio data, the above model is called A-VAE~\cite{sadeghiLAGH19}. The associated architecture is shown in Fig.~\ref{fig:vae_architectures}~(top).
	\subsection{Visual-only VAE}
	A visual VAE (V-VAE) is proposed in~\cite{sadeghiLAGH19}, assuming the same generative model as in \eqref{decoder_VAE} and \eqref{prior_VAE}. The 
	difference with the A-VAE is that, here, the posterior $ p(\mathbf{z}_n |\mathbf{s}_n) $ is approximated using visual-data only:
	\begin{equation}
		\label{eq:varappV}
		q(\mathbf{z}_n|\mathbf{v}_n,\bs{\phi}_v)= \mathcal{N}\Big(\bs{\mu}_z^v(\mathbf{v}_n), \text{diag}\Big(\bs{\sigma}_{z}^v(\mathbf{v}_n)\Big)\Big),
	\end{equation}
	where, $ \vb_n\in\Rbb^M $ is an embedding for the image of the
	speaker lips at frame $ n $, and $ \bs{\mu}_z^v(.): \mbb{R}^M \mapsto 
	\mbb{R}^L $ and $ \bs{\sigma}_{z}^v(.): \mbb{R}^M \mapsto 
	\mbb{R}_+^L $ denote neural networks with parameters $ \bs{\phi}_v $. Hence, V-VAE attempts to reconstruct clean speech using latent variables inferred from the lips images. The set of parameters, $ \lk\bs{\theta},\bs{\phi}_v\rk $, is estimated in the same way as A-VAE. Figure~\ref{fig:vae_architectures}~(middle) depicts the architecture of a V-VAE.
	\subsection{Audio-Visual VAE}
	An audio-visual VAE, called AV-VAE, is also presented in~\cite{sadeghiLAGH19} for speech modeling. The rationale of the AV-VAE  is to exploit the complementary between audio and visual modalities. The associated generative model is defined as:
	\begin{align}
		\label{decoder_AVVAE}
		\sbb_n | \mathbf{z}_n; \mathbf{v}_n&\sim \mathcal{N}_c\Big(\boldsymbol{0}, \text{diag}\Big(\bs{\sigma}_{s}(\mathbf{z}_n, \mathbf{v}_n)\Big)\Big), \\
		\label{prior_AVVAE}
		\mathbf{z}_n | \mathbf{v}_n&\sim \mathcal{N}\Big(\bs{\mu}_z^{av}(\mathbf{v}_n), \text{diag}\Big(\bs{\sigma}_{z}^{av}(\mathbf{v}_n)\Big)\Big), 
	\end{align}
	where, $ \bs{\sigma}_{s}(.,.): \mbb{R}^L\times \mbb{R}^M \mapsto 
	\mbb{R}_+^F $ is a neural network taking $ (\zb_n,\vb_n) $ as input. Furthermore, $ \bs{\mu}_z^{av}(.): \mbb{R}^M \mapsto 
	\mbb{R}^L $ and $ \bs{\sigma}_{z}^{av}(.): \mbb{R}^M \mapsto 
	\mbb{R}_+^L $ are neural networks parameterizing the mean and variance of the prior distribution of $ \zb_n $ using $ \vb_n $ as the input. Note that throughout the paper, $ \vb_n $ is treated as a deterministic piece of information, and so we do not model its generative process. The variational approximation to $ 
	p(\mathbf{z}_n |\mathbf{s}_n,\vb_n) $ takes a similar form as \eqref{eq:varapp}, except that $ \vb_n  $ is also fed to the associated neural network. The architecture of an AV-VAE is shown in Fig.~\ref{fig:vae_architectures}~(bottom).
	
	\section{The Mixture of Inference Networks VAE}\label{sec:prop}
	In this section, we aim to devise a framework able to choose the best combination between the auditory and visual modalities in the encoder, as opposed to systematically using both encodings through tight fusion, as done in AV-VAE. To achieve this goal, we propose a probabilistic mixture of an audio and a visual encoder, and name it mixture of inference networks VAE (MIN-VAE). Intuitively, the model learns to infer to which extent the approximate posterior of $\zb_n$ should be audio- or visual-based. The overall architecture is depicted in Fig.~\ref{fig:min_vae}. In the following, we introduce the mathematical formulations associated with the proposed MIN-VAE. The generative model is presented in Subsection~\ref{subsec:inf}, which uses a mixture of two different Gaussian distributions for the prior of the latent variables, as opposed to standard VAE and to the models in the previous section that use a standard Gaussian distribution. This innovative choice is motivated to ease the task of the generative model. Indeed, by modeling two different prior distributions, the decoder network can easily understand if the sample is audio-based or visual-based. Subsection~\ref{subsec:post} proposes a variational distribution to approximate the intractable posterior of the latent variables. This variational distribution is then used in Subsection~\ref{subsec:train} to derive the training algorithm for the overall MIN-VAE. Finally, noise modelling for speech enhancement at test time is discussed in Subsection~\ref{subsec:noise_model}.
	
	\subsection{The Generative Model}\label{subsec:inf}
	We assume that each latent code is generated either from an audio or from a visual prior. We model this with a mixing variable $\alpha_n\in\{0,1\}$ describing whether the latent code $\zb_n$ corresponds to the audio or to the visual prior. Once the latent code is generated from the corresponding prior, the speech frame $\sbb_n$ follows a complex Gaussian distribution with the variance computed by the decoder. We recall that the variance is a non-linear transformation of the latent code.
	\begin{figure*}[t!h!]
		\centering
		\rotatebox[origin=c]{90}{MIN-VAE}\hspace{8mm}\includegraphics[align=c,width=0.9\textwidth]{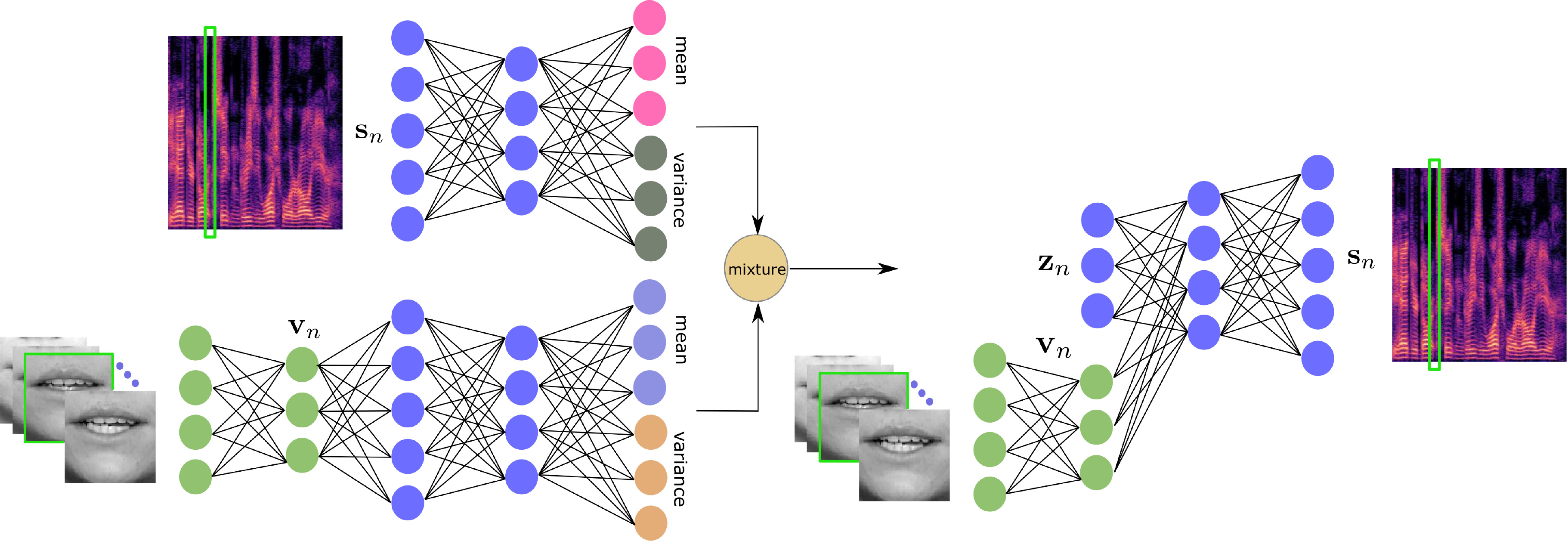}
		\caption{\label{fig:min_vae} Architecture of the proposed mixture of inference networks VAE (MIN-VAE). A mixture of an audio- and a visual-based encoder is used to approximate the intractable posterior distribution of the latent variables.}
	\end{figure*}
	
	Formally, each STFT time frame $ \sbb_n $ is modeled as:
	\begin{align}
		\label{decoder_VAE2}
		\sbb_n | \mathbf{z}_n; \vb_n &\sim \mathcal{N}_c\Big(\boldsymbol{0}, \text{diag}\Big(\bs{\sigma}_{s}(\mathbf{z}_n, \vb_n)\Big)\Big), \\
		\label{prior_VAE2}
		\zb_n|\alpha_n &\sim \Big[\Nc(\boldsymbol{\mu}_a, \sigma_a\Ib)\Big]^{\alpha_n}\cdot \Big[\Nc(\boldsymbol{\mu}_v, \sigma_v\Ib)\Big]^{1-\alpha_n}, \\
		\alpha_n &\sim \pi^{\alpha_n} \times (1-\pi)^{1-\alpha_n}\label{prior_alpha},
	\end{align}
	where the audio and visual priors are parametrized by $(\boldsymbol{\mu}_a, \sigma_a)$ and $(\boldsymbol{\mu}_v, \sigma_v)$ respectively, and $\alpha_n$ is assumed to follow a Bernoulli distribution with parameter $ \pi $. We propose two versions of this architecture, namely: MIN-VAE-v1 where the decoder \eqref{decoder_VAE2} takes the same form as \eqref{decoder_AVVAE} and uses explicitly visual information (see Fig.~\ref{fig:min_vae}), and MIN-VAE-v2 where the decoder \eqref{decoder_VAE2} takes the same form as \eqref{decoder_VAE} and does not use explicitly visual information. In both cases the parameters of the decoder are denoted by $\boldsymbol{\theta}$. The derivations will be done for the general case, that is MIN-VAE-v1.
	
	\subsection{The Posterior Distribution}\label{subsec:post}
	In order to estimate the parameters of the generative model described above, i.e. $ \bs{\psi}=\lk \boldsymbol{\mu}_a, \boldsymbol{\mu}_v, \sigma_a, \sigma_v\rk $, $ \bs{\theta} $, and $ \pi $, we follow a maximum likelihood procedure. To derive it, we need to compute the posterior of the latent variables:
	\begin{equation}
		p(\zb_n, \alpha_n|\sbb_n;\vb_n) = p(\zb_n|\sbb_n, \alpha_n;\vb_n)\cdot p(\alpha_n|\sbb_n;\vb_n).
		\label{eq:pz}
	\end{equation} 
	The individual factors in the right-hand side of the above equation cannot be computed in closed-form, due to the non-linear generative model. As similarly done in VAE, we pursue an amortized inference approach to approximate $ p(\zb_n|\sbb_n, \vb_n, \alpha_n) $ with a parametric Gaussian distribution defined as follows:
	\begin{equation}
		q(\zb_n|\sbb_n,\alpha_n;\vb_n,\bs{\phi}) = \begin{cases}
			q(\zb_n|\sbb_n;\bs{\phi}_a)& \alpha_n=1,\\
			q(\zb_n|\vb_n;\bs{\phi}_v)& \alpha_n=0,
		\end{cases}
	\end{equation}
	in which, $ \bs{\phi}=\lk\bs{\phi}_a, \bs{\phi}_v\rk $, and $ \bs{\phi}_a $ and $ \bs{\phi}_v $ denote the parameters of the associated audio and visual inference neural networks, taking the same architectures as those in \eqref{eq:varapp} and \eqref{eq:varappV}, respectively. For the posterior of $ \alpha_n $, i.e. $ p(\alpha_n|\sbb_n;\vb_n) $, we resort to a variational approximation, denoted $ r(\alpha_n) $. Put it all together, we have the following approximate posterior:
	\begin{equation}
		q(\zb_n|\sbb_n,\alpha_n;\vb_n,\bs{\phi})\cdot r(\alpha_n)\approx p(\zb_n, \alpha_n|\sbb_n;\vb_n).
		\label{eq:app_post}
	\end{equation}
	\subsection{Training the MIN-VAE}\label{subsec:train}
	In order to train the MIN-VAE, we devise an optimization procedure alternating between estimating $ \bs{\Theta}=\lk\bs{\phi}, \bs{\theta}, \bs{\psi}, \pi\rk $ and updating the variational posterior $r$. To do so, we first need to give an expression of the exact posterior distribution, so as to write the optimization function, namely the KL-divergence between the approximate variational posterior and the true posterior. To write the exact posterior, we will use the decomposition of the generative model in equations~(\ref{decoder_VAE2}) --~(\ref{prior_alpha}), and recall the definition $ \sbb=\lk \sbb_n\rk_{n=1}^{N_{tr}} $, and $\zb$, and define $\alphab$ and $\vb$ analogously. The full posterior of the latent variables writes:
	\begin{align}
		p(\zb,\alphab|\sbb;\vb,\bs{\theta}) &= \frac{p(\sbb,\zb,\alphab;\vb,\bs{\theta})}{p(\sbb;\vb,\bs{\theta})}\nonumber\\ &=\frac{p(\sbb|\zb;\vb,\bs{\theta})p(\zb|\alphab) p(\alphab)}{p(\sbb;\vb,\bs{\theta})}.
		\label{eq:newpost}
	\end{align}
	We then target the KL-divergence between the approximate posterior and the true posterior which reads:
	\begin{align}
		&{\cal D}_{KL}\Big(  q(\zb|\sbb,\alphab;\vb,\bs{\phi}) r(\alphab) \Big\| p(\zb,\alphab|\sbb;\vb,\bs{\theta}) \Big) =\nonumber \\
		& \sum_{\alphab}\int_{\mathbb{Z}} q(\zb|\sbb,\alphab;\vb,\bs{\phi}) r(\alphab) \log 
		\frac{q(\zb|\sbb,\alphab;\vb,\bs{\phi}) r(\alphab)p(\sbb;\vb,\bs{\theta})}{p(\sbb|\zb;\vb,\bs{\theta}) p(\zb|\alphab) p(\alphab) } \textrm{d}\zb  \nonumber\\
		&=
		-\mathcal{L}(\bs{\Theta}, r)+\log p(\sbb;\vb,\bs{\theta})\ge 0\label{eq:xavi-kl},
	\end{align}
	where 
	\begin{multline}
		\mathcal{L}(\bs{\Theta}, r) =\\ \sum_{\alphab}\int_{\mathbb{Z}} q(\zb|\sbb,\alphab;\vb,\bs{\phi}) r(\alphab) \log 
		\frac{p(\sbb|\zb;\vb,\bs{\theta}) p(\zb|\alphab) p(\alphab) }{q(\zb|\sbb,\alphab;\vb,\bs{\phi}) r(\alphab)} \textrm{d}\zb .
		\label{eq:elboA}
	\end{multline}
	From \eqref{eq:xavi-kl} we can see that $ \log p(\sbb;\vb,\bs{\theta})\ge \mathcal{L}(\bs{\Theta}, r)$. Therefore, instead of maximizing the intractable data log-likelihood $ \log p(\sbb;\vb,\bs{\theta}) $, we maximize its lower-bound, i.e. $\mathcal{L}(\bs{\Theta}, r) $, or equivalently:
	\begin{equation}
		\bs{\Theta}^*, r^* = \argmin_{\bs{\Theta}, r}-\mathcal{L}(\bs{\Theta}, r)
		\label{eq:lcost}
	\end{equation}
	subject to the constraint that $ r $ integrates to one. We solve this problem by alternately optimizing the cost over $ r $ and $ \bs{\Theta} $. In the following, the two optimization steps are discussed.
	
	\subsubsection{Optimizing w.r.t. $r(\alphab)$}
	With $ \bs{\Theta} $ being fixed to its current estimate, solving~(\ref{eq:lcost}) boils down to:
	\begin{equation}
		\min_{r} \sum_{\alpha_n} r_n(\alpha_n) \Big[ \log\frac{r_n(\alpha_n)}{p(\alpha_n)} + J_n(\alpha_n)\Big] 
		, \forall n,
	\end{equation}
	meaning that the optimal $r$ is separable on $n$, where,
	\begin{multline}
		J_n(\alpha_n) = \\\int_{\mathbb{Z}} q(\zb_n|\sbb_n,\alpha_n;\vb_n,\bs{\phi}) \log 
		\frac{q(\zb_n|\sbb_n,\alpha_n;\vb_n,\bs{\phi})}{p(\sbb_n|\zb_n;\vb_n,\bs{\theta})p(\zb_n|\alpha_n)}\textrm{d}\zb_n=\\{\cal D}_{KL}\Big(  q(\zb_n|\sbb_n,\alpha_n;\vb_n,\bs{\phi}) \Big\| p(\zb_n|\alpha_n) \Big)-\\\E_{q(\zb_n|\sbb_n,\alpha_n;\vb_n,\bs{\phi})}\Big[\log p(\sbb_n|\zb_n;\vb_n,\bs{\theta})\Big].
		\label{eq:j-aplha}
	\end{multline}
	Using calculus of variations, we find that $ r_n(\alpha_n) \propto p(\alpha_n) \exp\big(-J_n(\alpha_n)\big) $, which is a Bernoulli distribution. To find the associated parameter, we need to compute $ J_n(\alpha_n) $. Since the expectation involved in \eqref{eq:j-aplha} is intractable to compute, we approximate it using a single sample denoted $ \zb_n^{\alpha_n} $ drawn from $ q(\zb_n|\sbb_n,\alpha_n;\vb_n,\bs{\phi}) $, obtaining:
	\begin{align}
		&\tilde{J}_n(\alpha_n)=\nonumber\\
		&{\cal D}_{KL}\Big(  q(\zb_n|\sbb_n,\alpha_n;\vb_n,\bs{\phi}) \Big\| p(\zb_n|\alpha_n) \Big) - \log p(\sbb_n|\zb_n^{\alpha_n};\vb_n,\bs{\theta}),
		\label{eq:galph}
	\end{align}
	The parameter of the Bernoulli distribution then takes the following form:
	\begin{equation}
		\pi_n = g\Big( \tilde{J}_n(\alpha_n=0)-\tilde{J}_n(\alpha_n=1)+ \log \frac{\pi}{1-\pi}  \Big),
		\label{eq:pin-up}
	\end{equation}
	where $ g(x) = 1/(1+\exp(-x)) $ is the sigmoid function. Computations of the KL divergence terms are provided in Appendix~\ref{app:kl}.
	
	\subsubsection{Optimizing w.r.t. $ \bs{\Theta} $}
	With $ r $ being fixed to its current estimate, from \eqref{eq:lcost}, we can write the optimization over $ \bs{\Theta}$ as:
	\begin{align}
		&\min_{\bs{\Theta}}~\sum_{\alphab}\int_{\mathbb{Z}} q(\zb|\sbb,\alphab;\vb,\bs{\phi}) r(\alphab) \log 
		\frac{q(\zb|\sbb,\alphab;\vb,\bs{\phi}) r(\alphab)}{p(\sbb|\zb;\vb,\bs{\theta}) p(\zb|\alphab) p(\alphab) } \textrm{d}\zb \nonumber\\
		&=\min_{\bs{\Theta}}~\sum_{n=0}^{N_{tr}}\pi_n\Big({\cal D}_{KL}\Big(  q(\zb_n|\sbb_n;\bs{\phi}_a) \Big\| p(\zb_n|\alpha_n=1) \Big) -\nonumber\\&\E_{q(\zb_n|\sbb_n;\bs{\phi}_a)}\Big[\log p(\sbb_n|\zb_n;\vb_n,\bs{\theta})\Big]\Big)+\nonumber\\&(1-\pi_n)\Big({\cal D}_{KL}\Big(  q(\zb_n|\vb_n;\bs{\phi}_v) \Big\| p(\zb_n|\alpha_n=0) \Big) -\nonumber\\&\E_{q(\zb_n|\vb_n;\bs{\phi}_v)}\Big[\log p(\sbb_n|\zb_n;\vb_n,\bs{\theta})\Big]\Big)+{{\cal D}_{KL}}\Big(r\left(\alpha_n\right) \parallel p(\alpha_n)\Big).
		\label{eq:psisup}
	\end{align}
	As before, the expectations involved in the above equation are approximated with a single sample drawn from the associated posteriors.
	After computing the cost function, the parameters are updated using a re-parametrization trick along with a stochastic gradient descent algorithm, e.g. the Adam optimizer. Finally, optimizing \eqref{eq:psisup} over $ \pi $ leads to minimizing the following KL-divergence:
	\begin{equation}
		\label{eq:klalph}
		{{\cal D}_{KL}}\Big(r\left(\alpha_n\right) \parallel p(\alpha_n)\Big)=\pi_n\log\frac{\pi_n}{\pi}+(1-\pi_n)\log\frac{1-\pi_n}{\pi},
	\end{equation}
	yielding
	\begin{equation}
		\label{eq:pi-up}
		\pi=\frac{1}{N_{tr}}\sum_{n=1}^{N_{tr}}\pi_n.
	\end{equation}
	
	Now, with the derived variational inference formulas, we obtain the inference mixture for the MIN-VAE:
	\begin{align}
		p(\zb_n|\sbb_n;\vb_n) &=  \pi_n \,\mathcal{N}\Big(\bs{\mu}_z^a(\sbb_n), \text{diag}\Big(\bs{\sigma}_{z}^a(\mathbf{s}_n)\Big)\Big) \\
		&+ (1-\pi_n)\, \mathcal{N}\Big(\bs{\mu}_z^v(\mathbf{v}_n), \text{diag}\Big(\bs{\sigma}_{z}^v(\mathbf{v}_n)\Big)\Big).
		\label{eq:mixVAE}
	\end{align}
	The overall training algorithm then consists of alternating the variational distribution update of $ \alpha_n $ via \eqref{eq:pin-up}, the update of $ \bs{\phi} $, $ \bs{\theta} $, and $ \bs{\psi} $ via stochastic gradient descent of \eqref{eq:psisup}, and the update of $ \pi $ using \eqref{eq:pi-up}.
	
	\begin{figure*}[t]
		\centering
		\subfloat[SDR (dB)]{{\includegraphics[width=5.8cm]{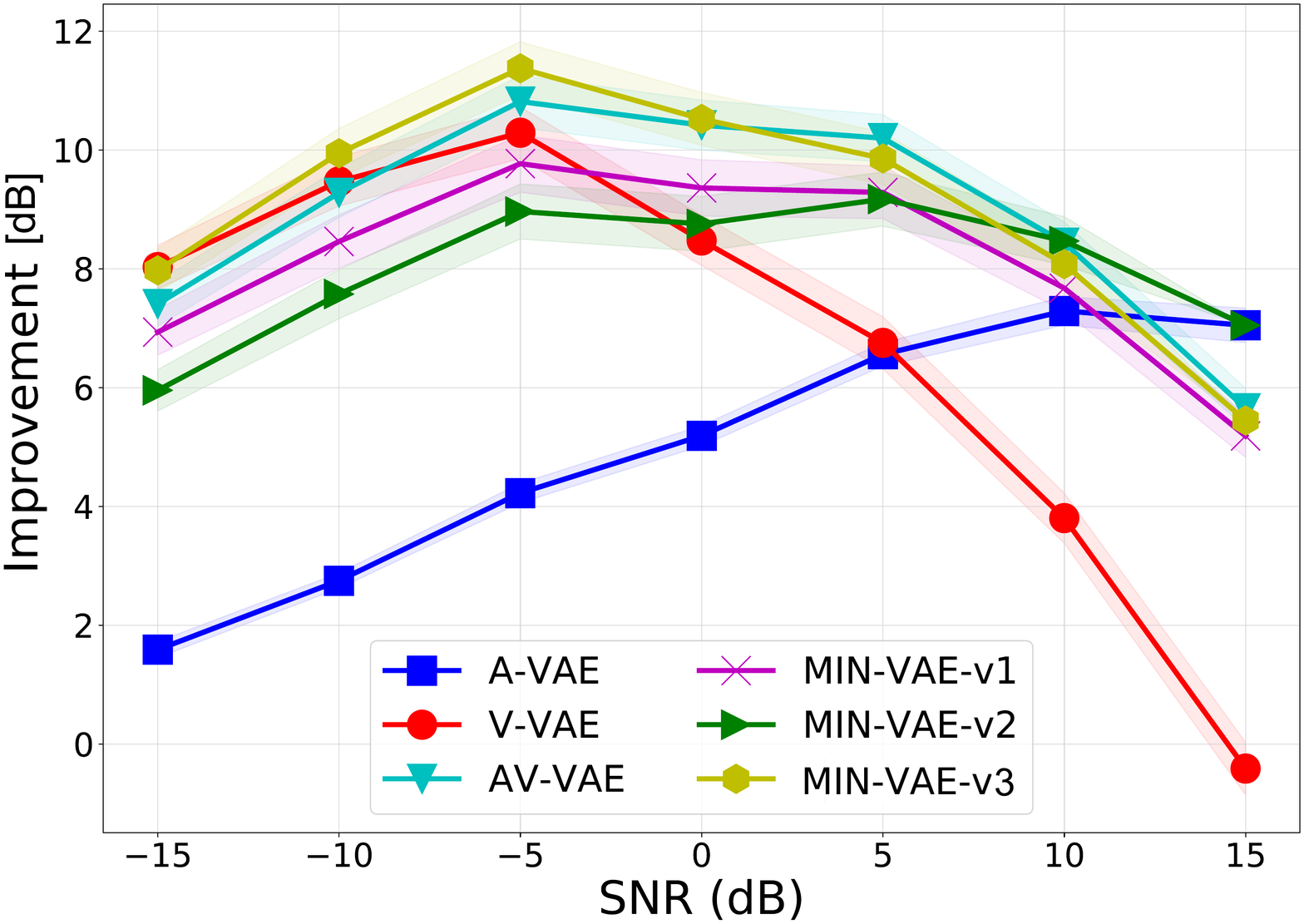} }}
		\subfloat[PESQ $ \lbrack -0.5,4.5 \rbrack$]{{\includegraphics[width=5.8cm]{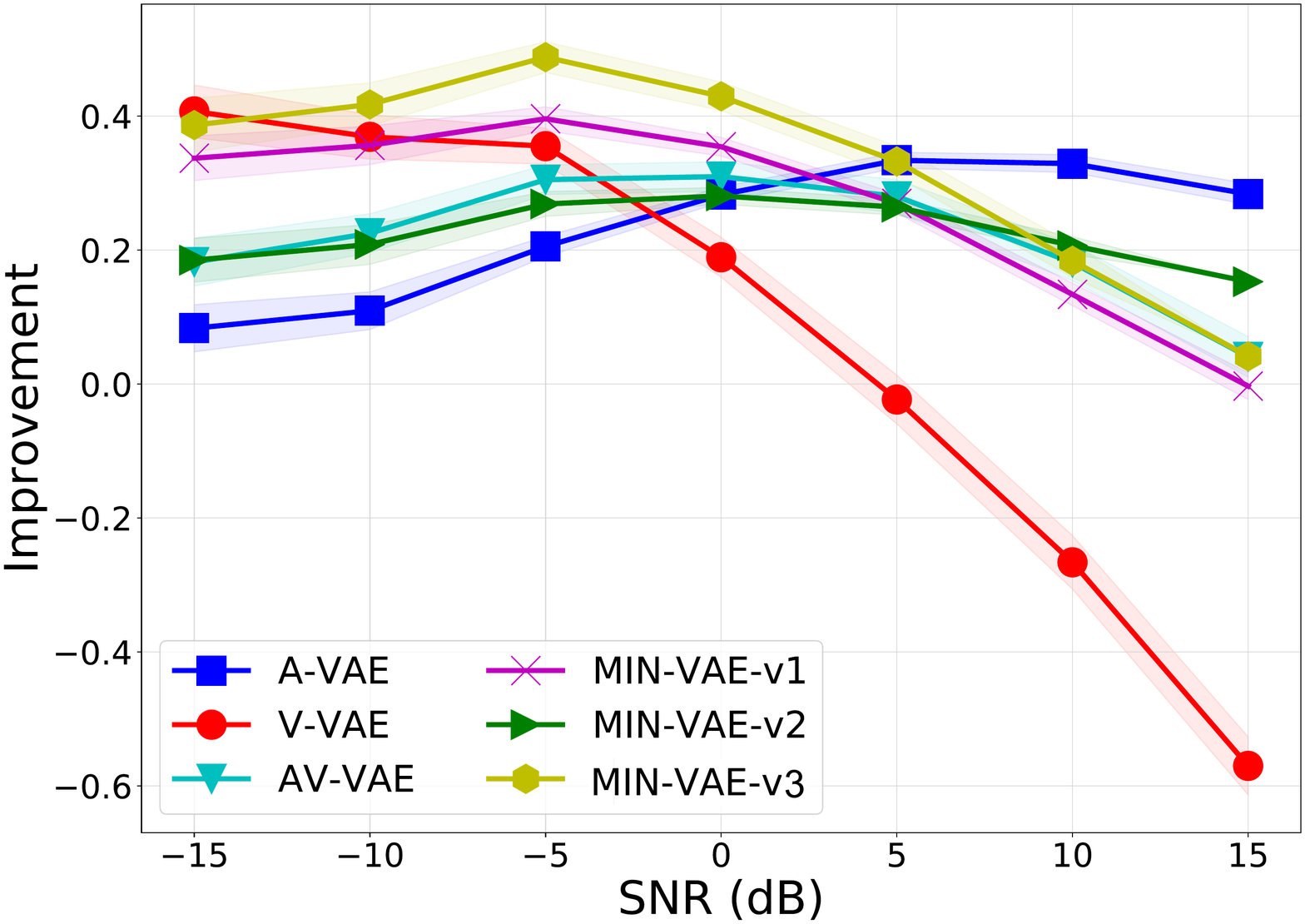} }}
		\subfloat[STOI $ \lbrack 0,1 \rbrack$]{{\includegraphics[width=5.8cm]{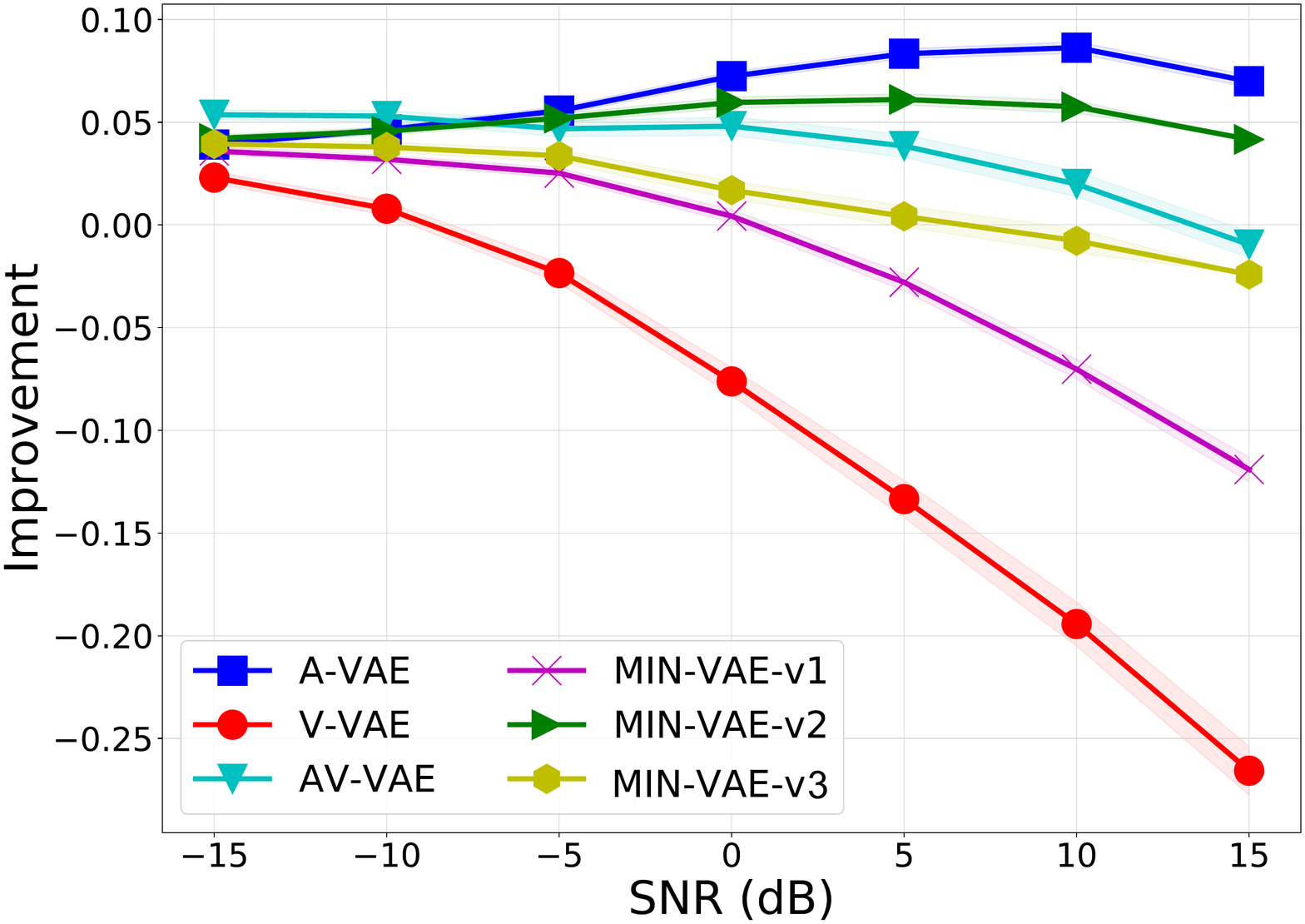} }}\\
		\subfloat[SDR (dB)]{{\includegraphics[width=5.8cm]{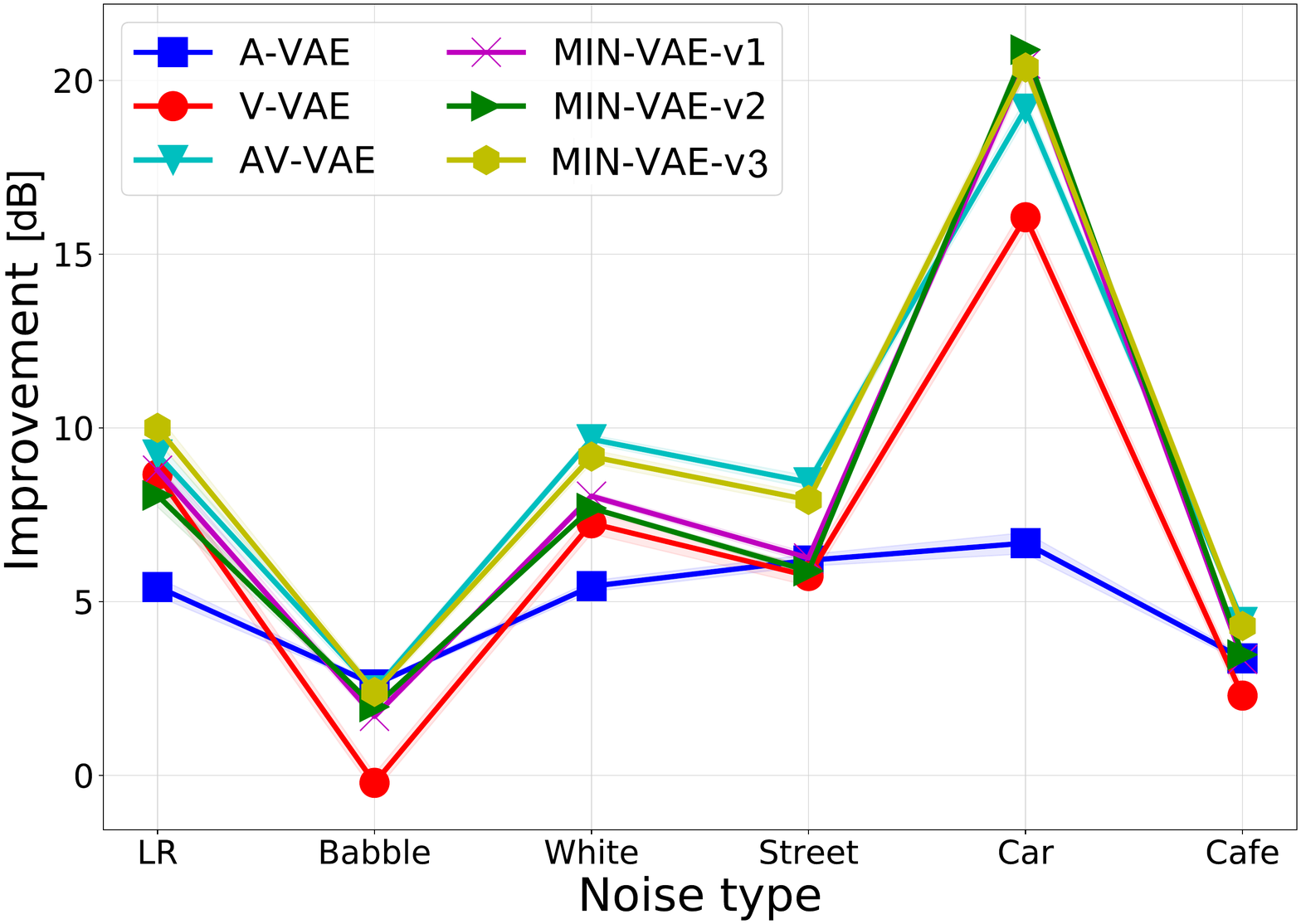} }}
		\subfloat[PESQ $ \lbrack -0.5,4.5 \rbrack$]{{\includegraphics[width=5.8cm]{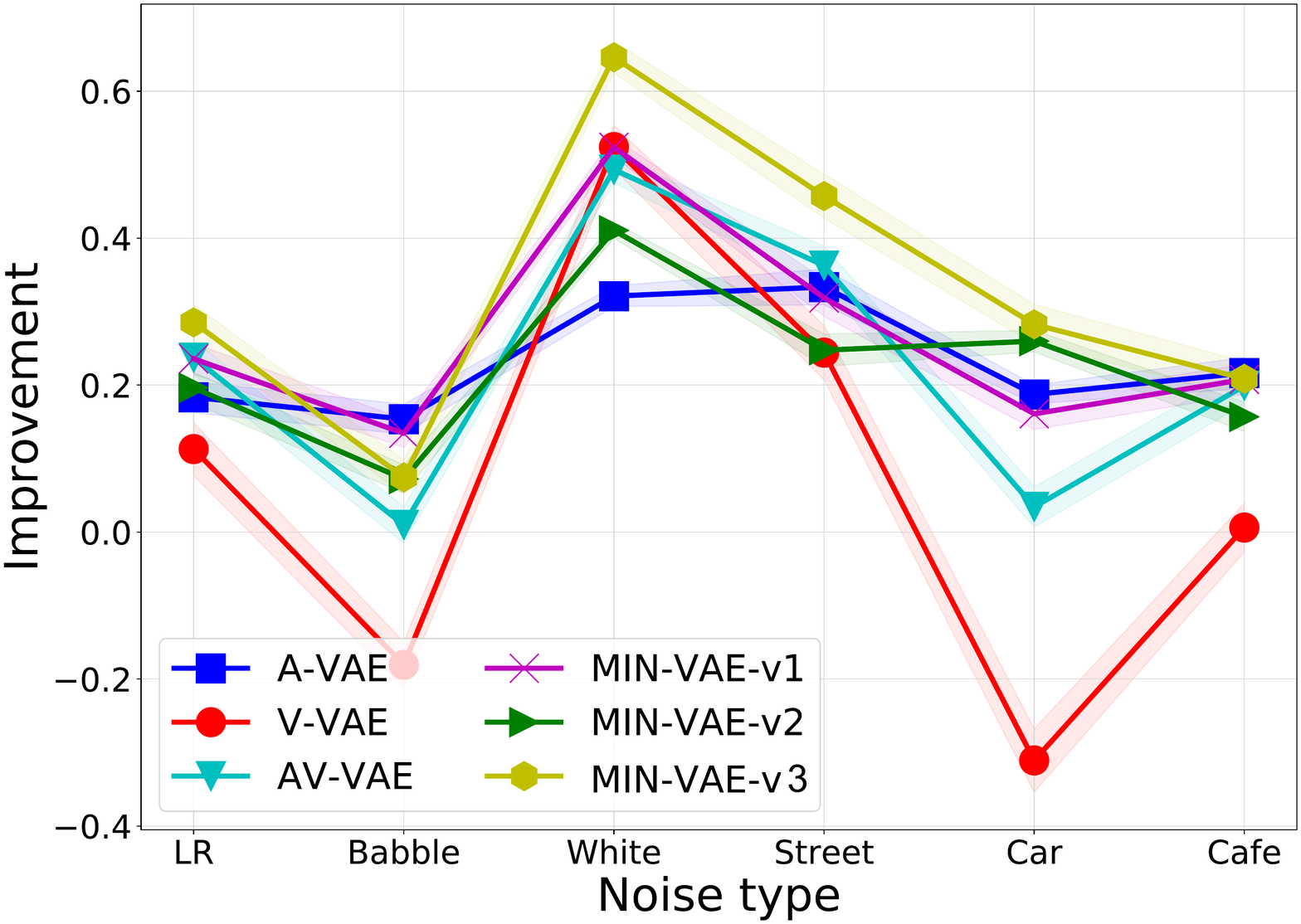} }}
		\subfloat[STOI $ \lbrack 0,1 \rbrack$]{{\includegraphics[width=5.8cm]{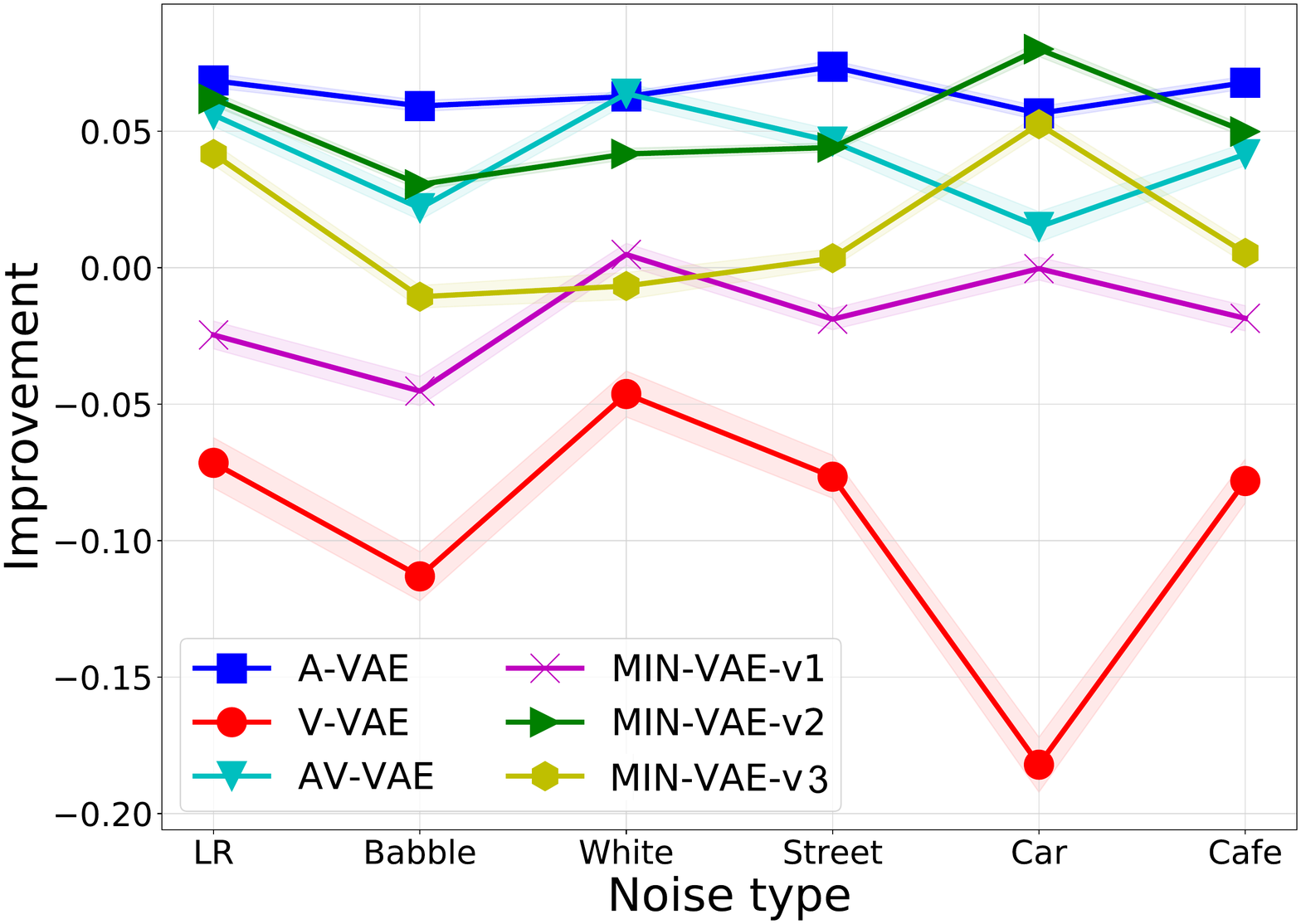} }}
		\caption{\label{fig:all_nonoise} Performance comparison of different VAE architectures for speech enhancement. Top row shows the averaged results in terms of input noise levels, whereas the bottom row reports the averaged results versus different noise types. Here, no noise was added to the input of the audio-encoders of MIN-VAE-v1 and MIN-VAE-v2 during training. }
	\end{figure*}
	
	\subsection{Noise Modeling}\label{subsec:noise_model}
	At test time, once the MIN-VAE is trained, the \ac{STFT} time frames of the observed noisy speech are modeled as $ \xb_n=\sbb_n+\bb_n $, for $ n=0,\ldots,N-1 $, with $ \bb_n $ denoting noise \ac{STFT} time frame. For the probabilistic modeling of $ \sbb_n $, we use the generative model trained on clean data (i.e.\ the previous section). For $ \bb_n $, the following \ac{NMF} based model is considered 
	\cite{Leglaive_MLSP18}:
	\begin{equation}
		\bb_n\sim \mathcal{N}\Big(\boldsymbol{0}, \text{diag}\Big(\Wb\bs{h}_n\Big)\Big),
		\label{eq:nmf}
	\end{equation}
	where, $ \Wb\in\Rbb_+^{F\times K} $, and $ \bs{h}_n $ denotes the $ n $-th column of $ \Hb\in\Rbb_+^{K\times N} $. The parameters, i.e. $ \lk \Wb, \Hb \rk $, as well as the unknown speech are then estimated following a variational inference method~\cite{bishop06}. This strategy is inspired by the recent literature~\cite{Leglaive_MLSP18,SadeA19a}. The details are provided in Appendix~\ref{app:se}.
	\section{Experiments}\label{sec:exp}
	In this section, we aim to evaluate the speech enhancement performance of different VAE architectures, including A-VAE~\cite{Leglaive_MLSP18}, V-VAE~\cite{sadeghiLAGH19}, AV-VAE~~\cite{sadeghiLAGH19}, and the proposed MIN-VAE.  To measure the performance, we use standard scores, including the \ac{SDR}~\cite{vincent2006performance}, the \ac{PESQ} ~\cite{rix2001perceptual}, and the \ac{STOI}~\cite{taal2011algorithm}. \ac{SDR} is measured in decibels (dB), while \ac{PESQ} and \ac{STOI} values lie in the intervals $[-0.5,4.5]$ and $ [0,1] $, respectively (the higher the better). For each measure, we report the averaged difference between the output value (evaluated on the enhanced speech signal) and the input value (evaluated on the noisy/unprocessed mixture signal). The average values of SDR, PESQ, and STOI computed on the input noisy speech signals are reported in Table~\ref{tab:input}. 
	\subsection{Experimental Set-up}
	\paragraph{Dataset} We use the NTCD-TIMIT dataset~\cite{Abde17}, which  contains \ac{AV} recordings from $ 56 $ English speakers with an Irish accent, uttering $ 5488 $ different TIMIT sentences~\cite{TIMIT}. The visual data consist of 30~FPS videos of lips \acp{ROI}. Each frame (\ac{ROI}) is of size 67$\times$67 pixels. The speech signal is sampled at $16$ kHz, and the audio spectral features are computed using an STFT window of 64~ms ($1024$ samples per frame) with 47.9\% overlap, hence $F=513$. The dataset is divided into $39$ speakers for training, $8$ speakers for validation, and $9$ speakers for testing, as proposed in~\cite{Abde17}. The test set includes about $ 1 $ hour noisy speech, along with their corresponding lips \ac{ROI}s, with six different noise types, including \textit{\ac{LR}}, \textit{White}, \textit{Cafe}, \textit{Car}, \textit{Babble}, and \textit{Street}, with noise levels: $ \lk -15,-10,-5,0,5,10,15 \rk $~dB. 
	\paragraph{Architecture and training details} The generative networks (decoders) of A-VAE and V-VAE consist of a single hidden layer with 128 nodes and hyperbolic tangent activations. The dimension of the latent space is $L=32$. The A-VAE encoder has 
	a single hidden layer with $ 128 $ nodes and hyperbolic tangent activations. The V-VAE encoder is similar to the A-VAE encoder, except for extracting visual features, embedding 
	lip \acp{ROI} into a feature vector $\mathbf{v}_n \in \mathbb{R}^M$, with $M=128$. This is composed of two fully-connected layers with $ 512 $ and $ 128 $ nodes. The dimension of the input corresponds to a single vectorized frame, namely $4489=67\times 67$. AV-VAE combines the architectures of A-VAE and V-VAE as illustrated in Fig.~\ref{fig:vae_architectures}. The audio 
	and the visual encoders in Fig.~\ref{fig:min_vae} share also the same architectures as those of A-VAE and V-VAE encoders, respectively. 
	
	To have a fair comparison, we fine-tunned the A-VAE and V-VAE of~\cite{sadeghiLAGH19}, which have been trained with a standard Gaussian prior for the latent variables, by using a parametric Gaussian prior, as the ones in \eqref{prior_VAE2}. The decoder parameters of MIN-VAE-v1 and MIN-VAE-v2 (see Section~\ref{subsec:inf}) are initialized with those of the pretrained AV-VAE and A-VAE, respectively. The parameters of the audio and the visual encoders are also initialized with the corresponding parameters in the pretrained A-VAE and V-VAE encoders. Then, all the parameters are fine-tuned using the Adam optimizer~\cite{kingma2014adam} with a step size of $10^{-4}$, for $ 100 $ epochs, and with a batch-size of $ 128 $. 
	
	We also considered another way to combine A-VAE with V-VAE, in which these two VAE architectures share the same decoder, and they are trained alternately. That is, at each epoch, the shared decoder is trained using latent samples coming from either the encoder of A-VAE or that of V-VAE. As a result, at each epoch, only the encoder parameters of the corresponding VAE, i.e. A-VAE or V-VAE, are updated while those of the other encoder are kept fixed. This training strategy is considered as a baseline where we do not use a mixture model for the encoder to automatically choose between the audio and the visual encoders. Instead, an alternating sampling from the two encoders is performed.  We refer to the resulting VAE as MIN-VAE-v3. A description of all the proposed VAE architectures is given in Table~\ref{tab:prop}.
	\begin{table}[ht]
		\centering
		\caption{Description of the proposed VAE networks.}
		\begin{tabular}[t]{ll}
			\hline
			\toprule
			Name&Description\\
			\hline
			\midrule
			{MIN-VAE-v1}&The architecture shown in Fig.~\ref{fig:min_vae}. \\
			\midrule
			\multirow{ 2}{*}{MIN-VAE-v2}&Same as MIN-VAE-v1 but without using visual modality\\ &in the decoder.\\
			\midrule
			\multirow{ 2}{*}{MIN-VAE-v3}&Alternately training an A-VAE and a V-VAE with a \\&shared decoder.\\
			\bottomrule
			\hline
		\end{tabular}
		\label{tab:prop}
	\end{table}
	
	\begin{table}[h]
		\centering
		\caption{Average score values computed on the input noisy speech signals.}
		\begin{tabular}{l|c|c|c|c|c|c|c}
			\hline
			\toprule
			\textbf{SNR (dB)} & \textbf{-15} & \textbf{-10} & \textbf{-5} & \textbf{0} & \textbf{5} & \textbf{10} & \textbf{15} \\ 
			\hline
			\midrule
			SDR (dB)          & -19          & -16          & -12.3       & -7.4       & -3         & 2           & 7           \\ 		\hline
			\midrule
			PESQ              & 1.22         & 1.31         & 1.44        & 1.69       & 1.98       & 2.32        & 2.64        \\ 		\hline
			\midrule
			STOI              & 0.32         & 0.38         & 0.46        & 0.55       & 0.65       & 0.76        & 0.81        \\ 		\bottomrule
			\hline
		\end{tabular}
		\label{tab:input}
	\end{table}
	
	\begin{figure*}[t]
		\centering
		\subfloat[SDR (dB)]{{\includegraphics[width=5.8cm]{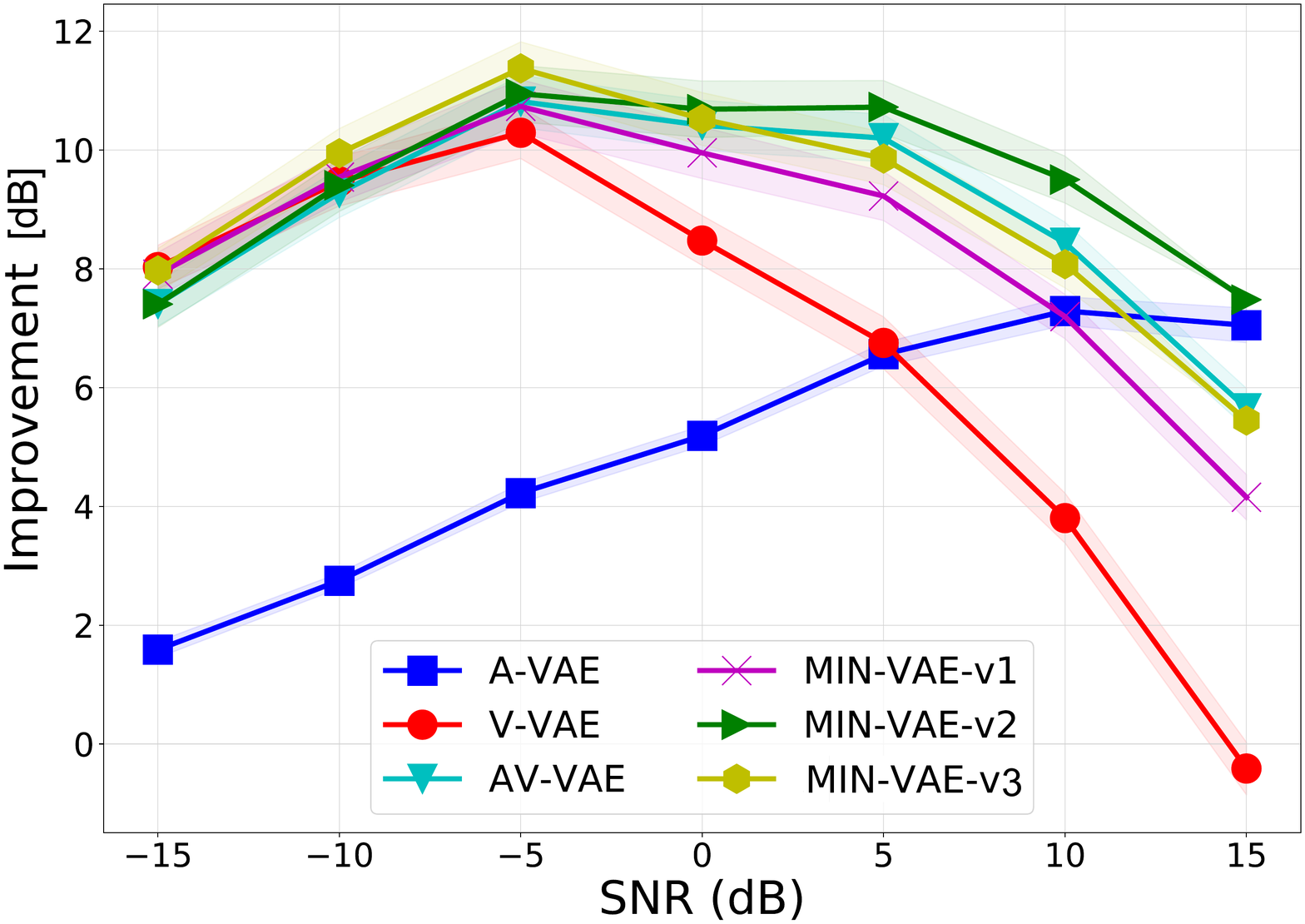} }}
		\subfloat[PESQ $ \lbrack -0.5,4.5 \rbrack$]{{\includegraphics[width=5.8cm]{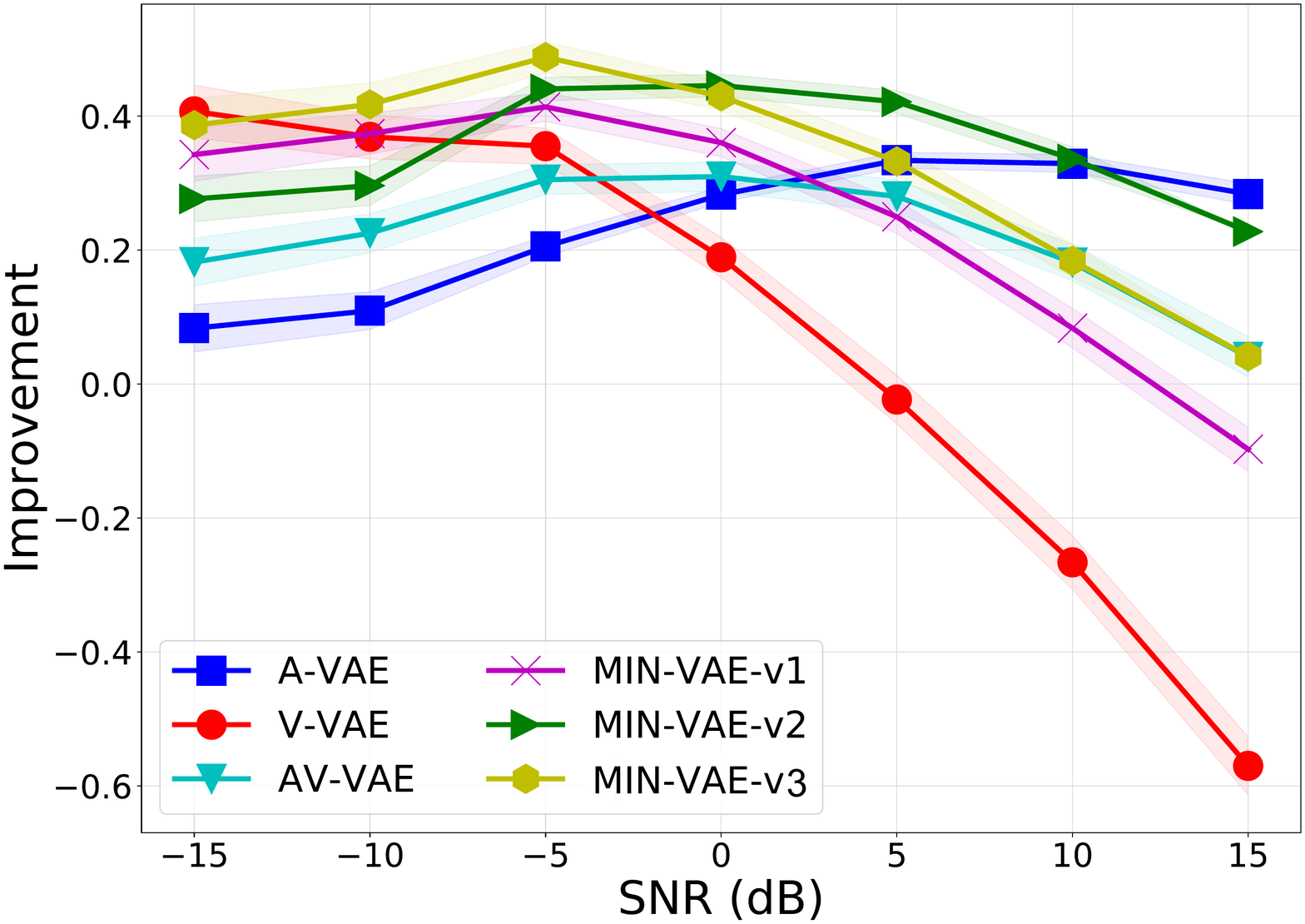} }}
		\subfloat[STOI $ \lbrack 0,1 \rbrack$]{{\includegraphics[width=5.8cm]{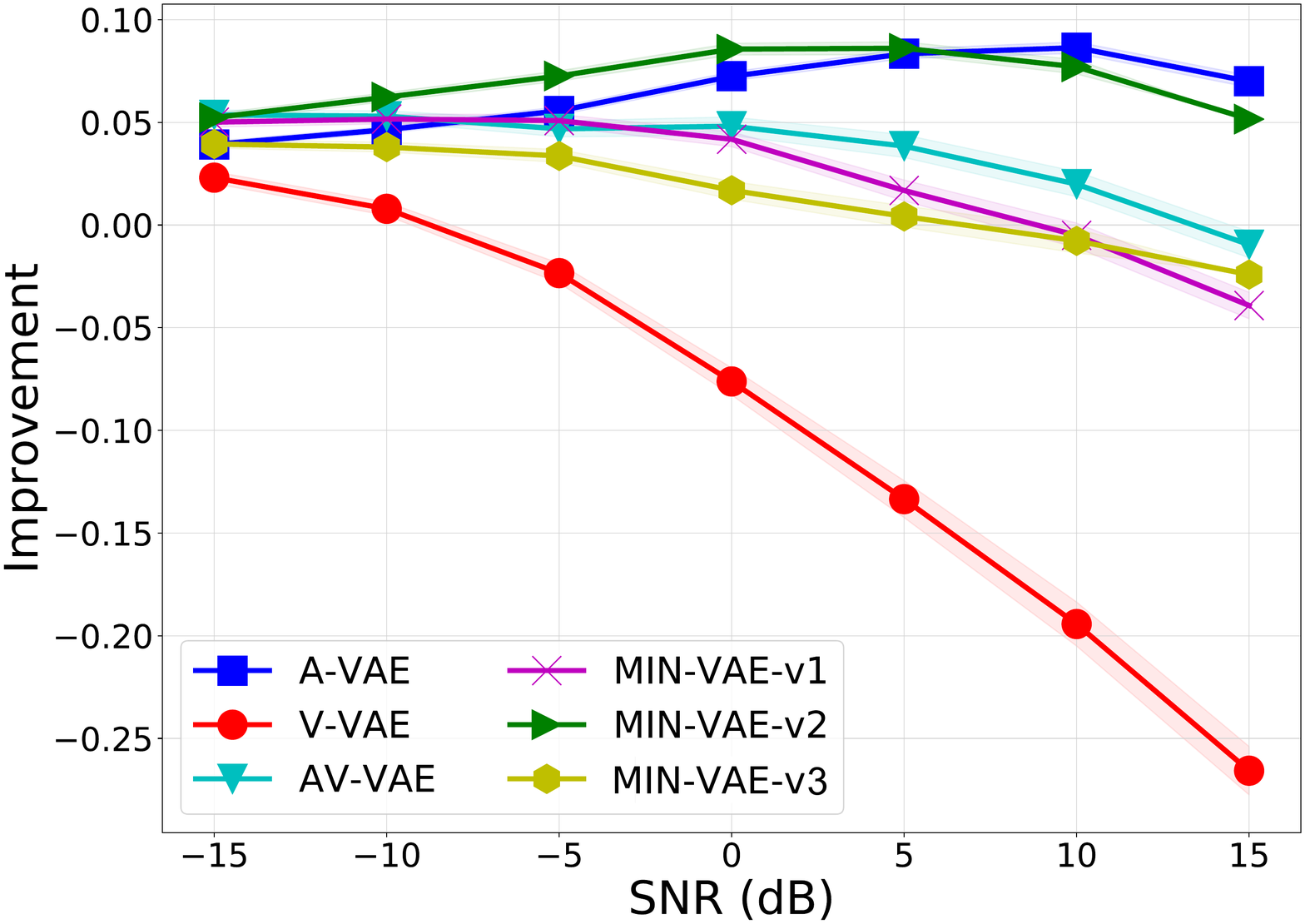} }}\\
		\subfloat[SDR (dB)]{{\includegraphics[width=5.8cm]{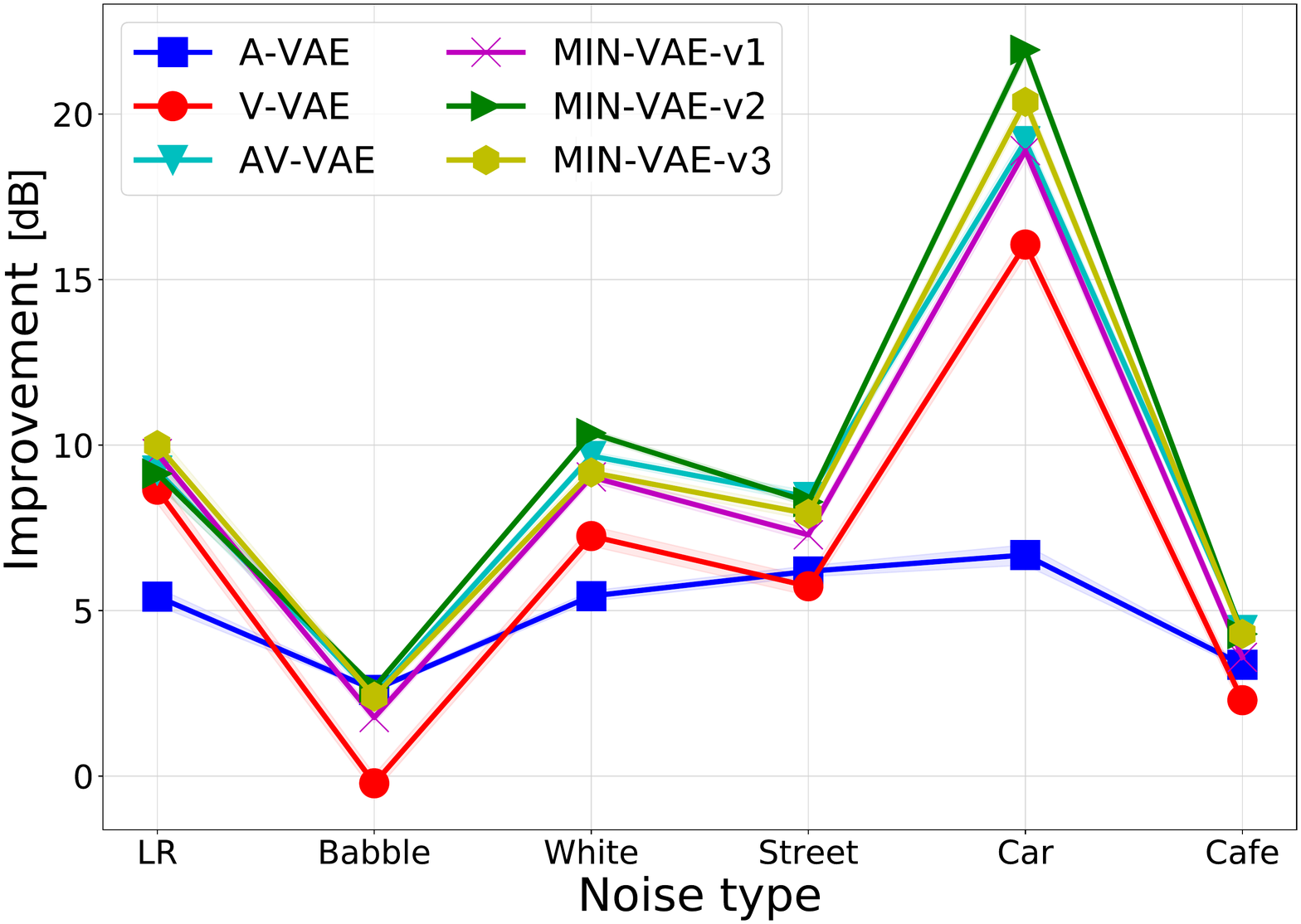} }}
		\subfloat[PESQ $ \lbrack -0.5,4.5 \rbrack$]{{\includegraphics[width=5.8cm]{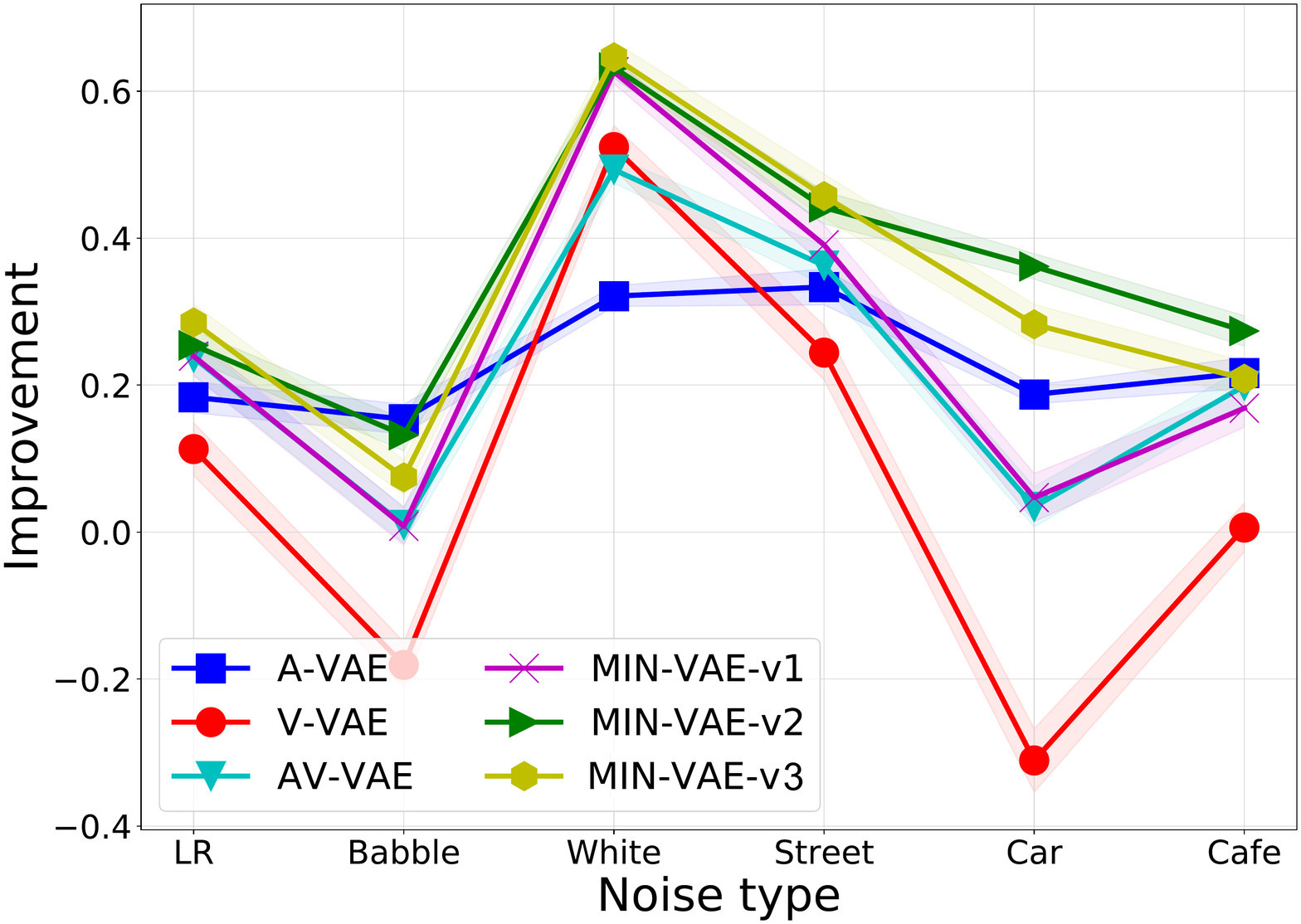} }}
		\subfloat[STOI $ \lbrack 0,1 \rbrack$]{{\includegraphics[width=5.8cm]{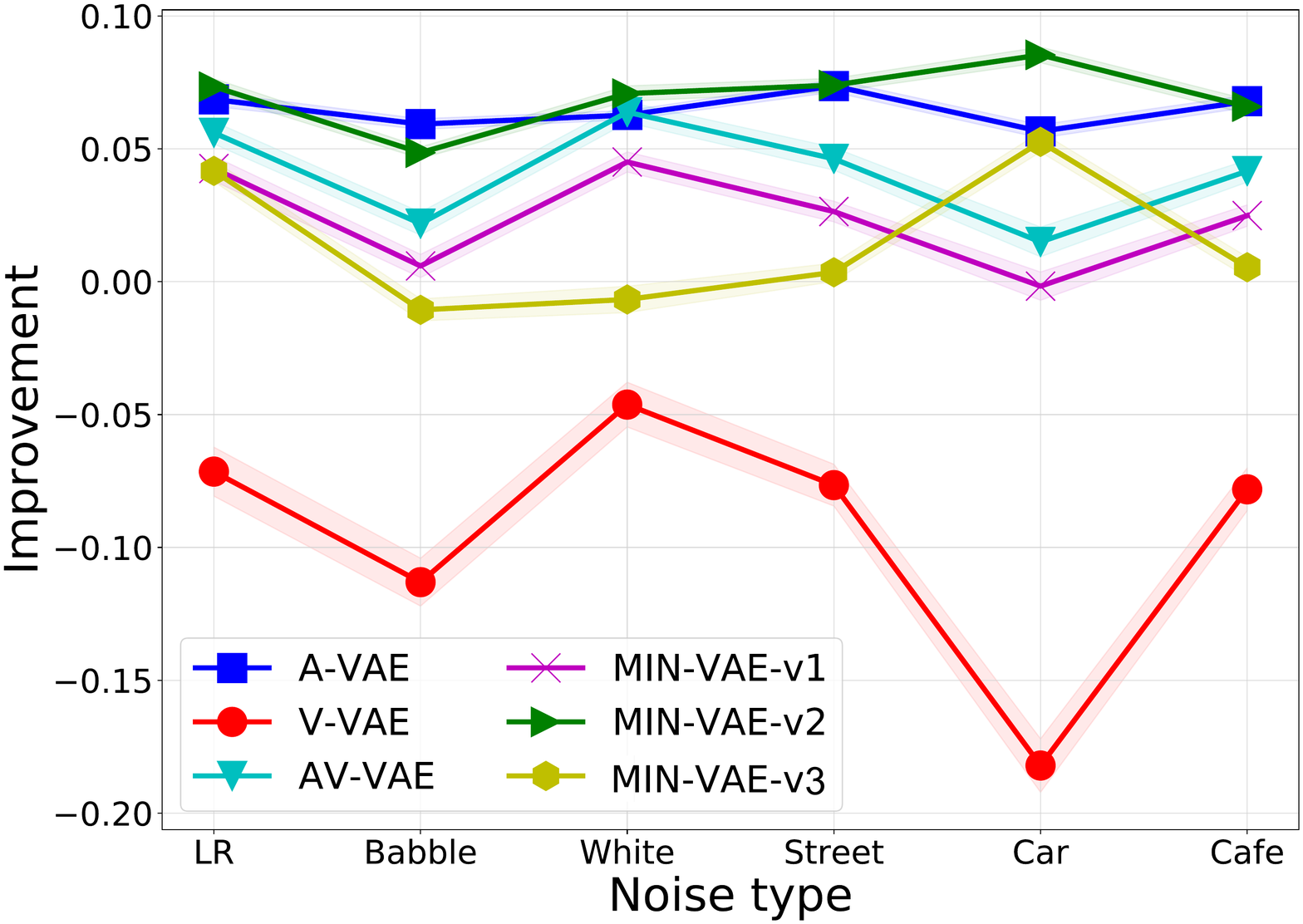} }}	
		\caption{\label{fig:all_noise} Performance comparison of different VAE architectures for speech enhancement. Top row shows the averaged results in terms of input noise levels, whereas the bottom row reports the averaged results versus different noise types. Here, some uniform noise was added to the input of the audio-encoders in MIN-VAE-v1 and MIN-VAE-v2 during training.}
	\end{figure*}
	
	\paragraph{Speech enhancement parameters} For all the methods, the rank of $ \Wb $ and $ \Hb $ in the noise model \eqref{eq:nmf} is set to $ K=10 $, and these matrices are randomly initialized with non-negative entries. At the first iteration of the inference algorithms, the Markov chain of the Metropolis-Hastings algorithm (see Section~\ref{sub:rz} in Appendix~B) is initialized by using the noisy observed speech and the visual features as input to the associated encoders, and taking the posterior mean as the initialization of the latent codes. For the proposed VAE architectures, i.e. MIN-VAE-v1, MIN-VAE-v2, and MIN-VAE-v3, the visual-encoders are used. 
	
	\subsection{Results and Discussion}
	Figure~\ref{fig:all_nonoise} summarizes the results of all the VAE architectures, in terms of SDR, PESQ, and STOI. The top row of this figure reports the averaged results versus different noise levels, whereas the bottom row shows the averaged results in terms of noise type. From this figure we can see that V-VAE performs pretty well at high noise levels. However, the intelligibility improvements in terms of STOI are not as good as those of the other algorithms. MIN-VAE-v3 outperforms other methods in terms of SDR and PESQ. Nevertheless, its intelligibility improvement is not satisfactory. The proposed MIN-VAE methods also outperform A-VAE, especially at high noise levels. As explained earlier, this might be due to the facts that the proposed networks efficiently make use of the robust initialization provided by the visual data, and also by the richer generative models (decoders) which are trained using both audio and visual latent codes. At high noise levels, MIN-VAE-v1 outperforms MIN-VAE-v2, implying the importance of using visual modality in the decoder when the input speech is very noisy. A related observation is that, MIN-VAE-v2 outperforms both MIN-VAE-v1 and AV-VAE when the level of noise is low, implying that the visual features in the generative model contribute mainly in high noise regimes. The average posterior probability of the $ \alpha_n $ variable given in \eqref{eq:pin-up} is $ 0.96 $, implying that the contribution of the audio encoder in generating the latent code is $ 96 $\%. Part of the worse performance of AV-VAE could be explained by the way the latent codes are initialized, which is based on concatenation of noisy audio and clean visual data. It is worth mentioning that in the low noise regime, the amount of performance improvement is decreasing for all the methods, as the speech signals are already clean enough. 
	
	{Regarding noise type, we see that the algorithms perform very differently. The \textit{Babble} noise is the most difficult noise environment according to the bottom row of Fig.~\ref{fig:all_nonoise}. In terms of SDR, all the methods show their best performance for the \textit{Car} noise, with a very large improvement achieved by the audio-visual based methods. In terms of PESQ, the \textit{White} noise is the easiest one for all the methods, especially MIN-VAE-v3 that shows the best performance. Finally, in terms of STOI, MIN-VAE-v2 achieves the best performance for the \textit{Car} noise.}
	
	To encourage the proposed MIN-VAE networks to make use of the visual data in the encoder more efficiently, we added some uniformly distributed noise, with the SNR being about 0 dB and fixed during training, to about one-third of speech spectrogram time frames that are fed to the audio encoders. {We also added noise to the audio encoder's inputs of A-VAE and AV-VAE. However, no performance improvements were observed.}\ The amount of noise and percentage of noisy frames have been found empirically. Figure~\ref{fig:all_noise} presents the results of this experiment. A clear performance improvement is observed compared to Fig.~\ref{fig:all_nonoise}, especially for MIN-AVE-v2. With this new training, the proposed algorithms outperform AV-VAE in all noise levels. The SDR improvements for high noise levels, however, are very close. {Regarding the improvement margin, we see that on average, MIN-VAE-v2 outperforms AV-VAE, about 1dB in terms of SDR (at high SNRs), more than 0.1 in terms of PESQ, and about 0.03-0.04 in terms of STOI. The PESQ and STOI improvements remain almost stable for different values of SNR. Overall, even if different MIN-VAE strategies may obtain the best improvement over AV-VAE depending on the measure and on the SNR, the superiority of MIN-VAE w.r.t.\ the AV-VAE is clear.}\ Finally, the best performing algorithm turns out to be MIN-VAE-v2, outperforming MIN-VAE-v3, especially at low levels of noise. The average posterior probability of the $ \alpha_n $ variable given in \eqref{eq:pin-up} is now $ 0.80 $, which is the best-performing value according to our experiments. Some audio examples are available at \url{https://team.inria.fr/perception/research/min-vae-se/}.
	
	%
	
	\section{Conclusions}
	Inspired by the importance of latent variable initialization for VAE-based speech enhancement, and as another way than simple concatenation to effectively fuse audio and visual modalities in the encoder of VAE, we proposed a mixture of inference (audio and visual encoders) networks, which are jointly trained with a shared generative network. The overall architecture is named MIN-VAE. A variational inference approach was proposed to estimate the parameters of the model. At test phase of the speech enhancement, the initialization of the latent variables, as required by the MCEM inference method, is based on the visual modality, which is assumed to be clean in contrast to audio data. As such, it provides a better performance than initializing with noisy audio data. This is confirmed by our experiments, comparing different VAE architectures.
	
{For future works, dynamical VAE architectures \cite{girin2020dynamical} will be investigated, which take into account the temporal correlation between audio and visual frames. This is expected to better handle visual modality, thus achieving superior performance compared to audio-only VAE. Furthermore, we will consider robustifying the proposed algorithms to noisy visual data, e.g. by using the mixture idea dveleoped in~\cite{SadeA21Switching}. Finally, reducing the computational complexity of the inference will be another promising research direction.}
	
	\appendices 
	\section{KL divergence computation}\label{app:kl}
	The KL divergence between two Gaussian distributions is given by the following lemma:
	\begin{lemma} Let $ p_1(\xb;\mub_1, \Sigb_1) $ and $ p_2(\xb;\mub_2, \Sigb_2) $ be two multivariate Gaussian distributions in $ \Rbb^n $. Then, the KL divergence between $ p_1 $ and $ p_2 $ reads:
		\begin{multline}
			{\cal D}_{KL}(p_1\| p_2 ) = \frac{1}{2} \Big( \log\frac{\det \Sigb_2}{\det \Sigb_1}-n+\trace(\Sigb_2^{-1}\Sigb_1)+\\(\mub_2-\mub_1)^T\Sigb_2^{-1}(\mub_2-\mub_1) \Big).
		\end{multline}\label{lem:kl}
	\end{lemma}
	
	Utilizing the above lemma, we can write the KL divergence term in \eqref{eq:galph} (for $\alpha_n=1$) as follows:
	\begin{multline}
		\label{eq:klcomp}
		\hspace*{-4mm}{{\cal D}_{KL}}\Big(q(\zb_n|\sbb_n;\bs{\phi}_a) \parallel p(\mathbf{z}_n|\alpha_n)\Big)=\frac{1}{2}\log\frac{\sigma_a^L}{\Big|\text{diag}\Big(\bs{\sigma}_{z}^a(\mathbf{s}_n)\Big)\Big|} -\\ 
		+\frac{\tr\Big(\text{diag}\Big(\bs{\sigma}_{z}^a(\mathbf{s}_n)\Big)\Big)+\|\bs{\mu}_z^a(\sbb_n)-\mub_a\|^2}{2\sigma_a}-\frac{L}{2},
	\end{multline}
	and analogously for the vision-based term ($\alpha_n=0$). 
	
	\section{Speech enhancement}\label{app:se}
	The generative model is given in \eqref{decoder_VAE2} -- \eqref{prior_alpha}, where all the parameters except $ \pi $ have already been trained on clean audio and visual data. The observations are noisy \ac{STFT} frames $ \xb=\lk \xb_n\rk_{n=0}^{N-1} $, as well as the visual data $ \vb=\lk \vb_n\rk_{n=0}^{N-1} $. The latent variables of the model are $ \sbb=\lk \sbb_n\rk_{n=0}^{N-1} $, $ \zb=\lk \zb_n\rk_{n=0}^{N-1} $, and $ \alphab=\lk \alpha_n\rk_{n=0}^{N-1} $. Furthermore, the parameters of the model are $ \Theta = \lk \mathbf{W},\mathbf{H}, \pi \rk $. 
	
	\subsection{Parameters Estimation}
	The full posterior is written as:
	\begin{multline}
		p(\sbb_n,\zb_n,\alpha_n|\xb_n;\vb_n,\Theta) \propto p(\xb_n,\sbb_n,\zb_n,\alpha_n;\vb_n,\Theta) = \\
		p(\xb_n|\sbb_n;\Theta)\times p(\sbb_n|\zb_n;\vb_n)\times p(\zb_n|\alpha_n) \times p(\alpha_n)
	\end{multline}
	To estimate the parameter set, we develop a variational inference method~\cite{bishop06}, where in the variational expectation step (VE-step), the above intractable posterior is approximated by a variational distribution $ r(\sbb_n, \zb_n, \alpha_n) $, as similarly done in~\cite{SadeA19a}. The maximization step (M-step) performs parameters update using the obtained variational distributions. We assume that $r$ factorizes as follows:
	\begin{equation}
		\label{eq:posterior}
		r(\sbb_n, \zb_n, \alpha_n) = r(\sbb_n) \times r(\zb_n)\times r(\alpha_n).
	\end{equation}
	where for notational convenience, we have omitted the dependence on $ \Theta $. Denoting the current estimate of the parameters as $ \Theta^{old} $, the VEM approach consists of iterating between the VE-steps and the M-step, which are detailed below.
	\subsubsection{VE-$r(\sbb_n)$ step}
	The variational distribution of $ \sbb_n $ is computed as~\cite{bishop06}:
	\begin{align}
		r(\sbb_n) \propto \exp\Big(& \E_{r(\zb_n)\cdot r(\alpha_n)}\Big[\log p(\xb_n,\sbb_n,\zb_n,\alpha_n;\vb_n,\Theta^{old})\Big]\Big)\nonumber\\
		= \exp\Big(&-\sum_f\Big[\frac{|x_{fn}-s_{fn}|^2}{(\Wb\Hb)_{fn}}+ \frac{|s_{fn}|^2}{\gamma_{fn}}\Big]\Big),\label{eq:rs}
	\end{align}
	where,
	\begin{equation}
		\label{eq:gamma_}
		\gamma_{fn}^{-1} = \E_{r(\zb_n)}\left[\frac{1}{{\sigma}_{s,f}(\mathbf{z}_n^{(d)}, \vb_n)}\right]\approx\frac{1}{D}\sum_{d=1}^{D} \frac{1}{{\sigma}_{s,f}(\mathbf{z}_n^{(d)}, \vb_n)},
	\end{equation}
	and $\lk\zb_n^{(d)}\rk_{d=1}^D$ is a sequence sampled from $ r(\zb_n) $. From \eqref{eq:rs}, we can see that $ r(s_{fn}) = \Nc_c(m_{fn},\nu_{fn}) $, where:
	\begin{equation}
		\label{eq:rs0}
		\begin{cases}
			m_{fn} & = \frac{\gamma_{fn}}{\gamma_{fn} + \left(\mathbf{W}\mathbf{H}\right)_{fn}}\cdot x_{fn} \\
			\nu_{fn} & = \frac{\gamma_{fn}\cdot \left(\mathbf{W}\mathbf{H}\right)_{fn}}{\gamma_{fn} + \left(\mathbf{W}\mathbf{H}\right)_{fn}}
		\end{cases}.
	\end{equation}
	\subsubsection{VE-$r(\zb_n)$ step}\label{sub:rz}
	The variational distribution of $ \zb_n $ can be computed by the following standard formula:
	\begin{align}
		r(\zb_n) &\propto \exp\Big( \E_{r(\sbb_n)\cdot r(\alpha_n)}\Big[\log p(\xb_n,\sbb_n,\zb_n,\alpha_n;\vb_n,\Theta^{old})\Big]\Big)\nonumber\\
		&\propto \exp\Big(\sum_f -\log\Big({{\sigma}_{s,f}(\mathbf{z}_n, \vb_n)} \Big)
		-\frac{|m_{fn}|^2+\nu_{fn}}{{\sigma}_{s,f}(\mathbf{z}_n, \vb_n)}\nonumber\\&+\sum_{\alpha_n\in\lk 0,1\rk}r(\alpha_n)\cdot \Big[\log p(\zb_n|\alpha_n)\Big]\Big)\triangleq \tilde{r}(\zb_n)
	\end{align}
	This gives us an unnormalized distribution $ \tilde{r}(\zb_n) $ whose normalization constant cannot be computed in closed-from, due to the non-linear terms. However, we use the Metropolis-Hastings algorithm~\cite{bishop06} to sample from it. To that end, we need to start with an initialization, $ \zb^{(0)} $. At the beginning of the inference, $ \zb^{(0)} $ is set to be the posterior mean in the output of the visual-encoder, i.e. the bottom-left network in Fig.~\ref{fig:min_vae}, where $ \vb_n $ is given as the input. Then, a candidate sample denoted $ \zb^{(c)} $ is obtained by sampling from a proposal distribution, usually chosen to be a Gaussian:
	\begin{equation}
		\zb^{(c)}|\zb^{(0)}\sim \Nc(\zb^{(0)}, \epsilon\Ib),
	\end{equation}
	where, $ \epsilon>0 $ controls the speed of convergence. Then, $ \zb^{(c)} $ is set to be the next sample $ \zb^{(1)} $ with the following probability:
	\begin{equation}
		p=\min\Big(1,\frac{\tilde{r}(\zb^{(c)})}{\tilde{r}(\zb^{(0)})}\Big).
	\end{equation}
	That means, some $ u $ is drawn from a uniform distribution between $ 0 $ and $ 1 $. Then, if $ u< p $, the sample is accepted and $ \zb^{(1)}=\zb^{(c)} $. Otherwise, it is rejected and $ \zb^{(1)}=\zb^{(0)} $. This procedure is repeated until the required number of samples is achieved. The first few samples are usually discarded, as they are not so reliable.
	\subsubsection{VE-$r(\alpha_n)$ step}
	The variational distribution of $ \alpha_n $ is computed as:
	\begin{align}
		r(\alpha_n)&\propto \exp\Big(\E_{r(\sbb_n)\cdot r(\zb_n)}\Big[\log p(\xb_n,\sbb_n,\zb_n,\alpha_n;\vb_n,\Theta^{old})\Big]\Big)\nonumber\\
		&\propto p(\alpha_n)\times\exp\Big(\E_{r(\zb_n)}\Big[\alpha_n\cdot\log p(\zb_n|\alpha_n=1)\nonumber\\&+(1-\alpha_n)\cdot\log p(\zb_n|\alpha_n=0)\Big]\Big)
	\end{align}
	which is a Bernoulli distribution with the following parameter:
	\begin{equation}
		\pi_n = g\Big(\E_{r(\zb_n)} \Big[\log\frac{p(\zb_n|\alpha_n=1)}{p(\zb_n|\alpha_n=0)}\Big] + \log \frac{\pi}{1-\pi}  \Big),
	\end{equation}
	with $ g(.) $ being the sigmoid function.
	\subsubsection{M-step}
	After updating all the variational distributions, the next step is to update the set of parameters, i.e. $ \Theta = \lk \mathbf{W},\mathbf{H}, \pi \rk $. To do so, we need to optimize the complete-data log-likelihood which reads:
	\begin{align}
		Q(\Theta;\Theta^{old})&=\E_{r(\sbb)\cdot r(\zb)\cdot r(\alphab)}\Big[\log p(\xb,\sbb,\zb,\alphab;\vb,\Theta)\Big]\nonumber\\
		&\stackrel{cte.}{=} \sum_{f,n}- \frac{|x_{fn}-m_{fn}|^2+\nu_{fn}}{(\Wb\Hb)_{fn}}-\log (\Wb\Hb)_{fn}\nonumber\\
		&\qquad+\pi_n\log \pi + (1-\pi_n)\log (1-\pi)
	\end{align}
	The update formulas for $ \mathbf{W} $ and $ \mathbf{H} $ can be obtained by using standard multiplicative updates~\cite{FevotBD09}:
	\begin{equation}
		\mathbf{H} \leftarrow \mathbf{H} \odot  \frac{\mathbf{W}^\top \left( \Vb \odot
			\left(\mathbf{W} \mathbf{H}\right)^{\odot-2} \right)}{\mathbf{W}^\top \left(\mathbf{W} \mathbf{H}\right)^{\odot-1} },
		\label{updateH}
	\end{equation}
	\begin{equation}
		\mathbf{W} \leftarrow \mathbf{W} \odot \frac{ \left( \Vb \odot
			\left(\mathbf{W} \mathbf{H}\right)^{\odot-2} \right)\mathbf{H}^\top}{\left(\mathbf{W} \mathbf{H}\right)^{\odot-1}\mathbf{H}^\top  },
		\label{updateW}
	\end{equation}
	where $ \Vb = \Big[|x_{fn}-m_{fn}|^2+\nu_{fn}\Big]_{(f,n)} $, and $\odot$ denotes element-wise operation. Optimizing over $ \pi $ leads to a similar update formula as in \eqref{eq:pi-up}: 
	\begin{equation}
		\pi=\frac{1}{N}\sum_{n=1}^{N}\pi_n.
	\end{equation}
	\subsection{Speech Estimation}
	Let $ \Theta^* = \lk \mathbf{W}^*,\mathbf{H}^*, \pi^* \rk $ denote the optimal set of parameters found by the above VEM procedure. An estimation of the clean speech is then obtained as the variational posterior mean ($ \forall f, n $):
	\begin{equation}
		\hat{s}_{fn}=\E_{r(s_{fn})}[s_{fn}]=\frac{\gamma_{fn}^*}{\gamma_{fn}^* + \left(\mathbf{W}^*\mathbf{H}^*\right)_{fn}}\cdot x_{fn},
	\end{equation}
	where, $ \gamma_{fn}^* $, defined in \eqref{eq:gamma_}, is computed using the optimal parameters.
	\bibliographystyle{IEEEbib}
	\bibliography{myref_compressed}

\begin{thebibliography}{10}

\bibitem{VincVG18}
E.~Vincent, T.~Virtanen, and S.~Gannot,
\newblock {\em Audio Source Separation and Speech Enhancement},
\newblock Wiley, 2018.

\bibitem{loizou2007speech}
P.~C. Loizou,
\newblock {\em Speech enhancement: theory and practice},
\newblock CRC press, 2007.

\bibitem{boll1979suppression}
S.~Boll,
\newblock ``Suppression of acoustic noise in speech using spectral
  subtraction,''
\newblock {\em IEEE Transactions on Acoustics, Speech, and Signal Processing},
  vol. 27, no. 2, pp. 113--120, 1979.

\bibitem{lim1979enhancement}
J.~S. Lim and A.~V. Oppenheim,
\newblock ``Enhancement and bandwidth compression of noisy speech,''
\newblock {\em Proceedings of the IEEE}, vol. 67, no. 12, pp. 1586--1604, 1979.

\bibitem{wang2018supervised}
W.~DeLiang and J.~Chen,
\newblock ``Supervised speech separation based on deep learning: An overview,''
\newblock {\em IEEE Transactions on Audio, Speech, and Language Processing},
  vol. 26, no. 10, pp. 1702--1726, 2018.

\bibitem{xu2015regression}
Y.~Xu, J.~Du, L.-R. Dai, and C.-H. Lee,
\newblock ``A regression approach to speech enhancement based on deep neural
  networks,''
\newblock {\em IEEE Transactions on Audio, Speech, and Language Processing},
  vol. 23, no. 1, pp. 7--19, 2015.

\bibitem{li:hal-02264247}
X.~Li and R.~Horaud,
\newblock ``Multichannel speech enhancement based on time-frequency masking
  using subband long short-term memory,''
\newblock in {\em IEEE Workshop on Applications of Signal Processing to Audio
  and Acoustics}, 2019, pp. 1--5.

\bibitem{wilson2008speech}
K.~Wilson, B.~Raj, P.~Smaragdis, and A.~Divakaran,
\newblock ``Speech denoising using nonnegative matrix factorization with
  priors,''
\newblock in {\em Proc. IEEE International Conference on Acoustics, Speech and
  Signal Processing (ICASSP)}, Las Vegas, USA, 2008, pp. 4029--4032.

\bibitem{mohammadiha2013supervised}
N.~Mohammadiha, P.~Smaragdis, and A.~Leijon,
\newblock ``Supervised and unsupervised speech enhancement using nonnegative
  matrix factorization,''
\newblock {\em IEEE Transactions on Audio, Speech, and Language Processing},
  vol. 21, no. 10, pp. 2140--2151, 2013.

\bibitem{SediBRJ17}
F.~Sedighin, M.~Babaie-Zadeh, B.~Rivet, and C.~Jutten,
\newblock ``Multimodal soft nonnegative matrix co-factorization for convolutive
  source separation,''
\newblock vol. 65, no. 12, pp. 3179--3190, 2017.

\bibitem{ISNMF}
C.~F{\'e}votte, N.~Bertin, and J.-L. Durrieu,
\newblock ``{Nonnegative matrix factorization with the Itakura-Saito
  divergence: With application to music analysis},''
\newblock {\em Neural computation}, vol. 21, no. 3, pp. 793--830, 2009.

\bibitem{SmarRS07_nmf}
F.~Smaragdis, B.~Raj, and M.~Shashanka,
\newblock ``Supervised and semi-supervised separation of sounds from
  single-channel mixtures,''
\newblock in {\em Proc. Int. Conf. Indep. Component Analysis and Signal
  Separation}, 2007, pp. 414--421.

\bibitem{MysoS11}
G.~J.~Mysore and P.~Smaragdis,
\newblock ``A non-negative approach to semi-supervised separation of speech
  from noise with the use of temporal dynamics,''
\newblock in {\em 2011 IEEE International Conference on Acoustics, Speech and
  Signal Processing (ICASSP)}, 2011.

\bibitem{SunZZ16}
M.~Sun, X.~Zhang, and T.~F. Zheng,
\newblock ``Unseen noise estimation using separable deep auto encoder for
  speech enhancement,''
\newblock vol. 24, no. 1, pp. 93--104, 2016.

\bibitem{bando2018statistical}
Y.~Bando, M.~Mimura, K.~Itoyama, K.~Yoshii, and T.~Kawahara,
\newblock ``Statistical speech enhancement based on probabilistic integration
  of variational autoencoder and non-negative matrix factorization,''
\newblock in {\em Proc. IEEE International Conference on Acoustics, Speech, and
  Signal Processing (ICASSP)}, 2018, pp. 716--720.

\bibitem{Leglaive_MLSP18}
S.~Leglaive, L.~Girin, and R.~Horaud,
\newblock ``A variance modeling framework based on variational autoencoders for
  speech enhancement,''
\newblock in {\em Proc. IEEE International Workshop on Machine Learning for
  Signal Processing (MLSP)}, 2018, pp. 1--6.

\bibitem{SekiguchiAPSIPA2018}
K.~Sekiguchi, Y.~Bando, K.~Yoshii, and T.~Kawahara,
\newblock ``Bayesian multichannel speech enhancement with a deep speech
  prior,''
\newblock in {\em Proc. Asia-Pacific Signal and Information Processing
  Association Annual Summit and Conference (APSIPA ASC)}, 2018, pp. 1233--1239.

\bibitem{PariDV19}
M.~Pariente, A.~Deleforge, and E.~Vincent,
\newblock ``A statistically principled and computationally efficient approach
  to speech enhancement using variational autoencoders,''
\newblock in {\em Proc. Conference of the International Speech Communication
  Association (INTERSPEECH)}, 2019.

\bibitem{Leglaive_ICASSP2019b}
S.~Leglaive, U.~\c{S}im\c{s}ekli, A.~Liutkus, L.~Girin, and R.~Horaud,
\newblock ``Speech enhancement with variational autoencoders and alpha-stable
  distributions,''
\newblock in {\em Proc. IEEE International Conference on Acoustics, Speech, and
  Signal Processing (ICASSP)}, 2019, pp. 541--545.

\bibitem{KameLIM19}
H.~Kameoka, L.~Li, S.~Inoue, and S.~Makino,
\newblock ``Supervised determined source separation with multichannel
  variational autoencoder,''
\newblock {\em Neural Computation}, vol. 31, no. 9, pp. 1--24, 2019.

\bibitem{nguyen2020deep}
V.~Nhat~Nguyen, M.~Sadeghi, E.~Ricci, and X.~Alameda-Pineda,
\newblock ``Deep variational generative models for audio-visual speech
  separation,''
\newblock {\em arXiv preprint arXiv:2008.07191}, 2020.

\bibitem{RezeMW14}
D.~J. Rezende, S.~Mohamed, and D.~Wierstra,
\newblock ``Stochastic backpropagation and approximate inference in deep
  generative models,''
\newblock in {\em Proceedings of the 31st International Conference on Machine
  Learning (ICML)}, 2014.

\bibitem{KingW14}
D.~P. Kingma and M.~Welling,
\newblock ``Auto-encoding variational bayes,''
\newblock in {\em International Conference on Learning Representations (ICLR)},
  2014.

\bibitem{cech2013active}
Jan Cech, Ravi Mittal, Antoine Deleforge, Jordi Sanchez-Riera, Xavier
  Alameda-Pineda, and Radu Horaud,
\newblock ``Active-speaker detection and localization with microphones and
  cameras embedded into a robotic head,''
\newblock in {\em IEEE-RAS International Conference on Humanoid Robots}, 2013,
  pp. 203--210.

\bibitem{ban2019variational}
Yutong Ban, Xavier Alameda-Pineda, Laurent Girin, and Radu Horaud,
\newblock ``Variational bayesian inference for audio-visual tracking of
  multiple speakers,''
\newblock {\em IEEE transactions on pattern analysis and machine intelligence},
  2019.

\bibitem{ban2017tracking}
Yutong Ban, Xavier Alameda-Pineda, Fabien Badeig, Sileye Ba, and Radu Horaud,
\newblock ``Tracking a varying number of people with a visually-controlled
  robotic head,''
\newblock in {\em IEEE/RSJ International Conference on Intelligent Robots and
  Systems}, 2017, pp. 4144--4151.

\bibitem{girin2001audio}
L.~Girin, J.-L. Schwartz, and G.~Feng,
\newblock ``Audio-visual enhancement of speech in noise,''
\newblock {\em The Journal of the Acoustical Society of America}, vol. 109, no.
  6, pp. 3007--3020, 2001.

\bibitem{AfouCZ18}
T.~Afouras, J.~S. Chung, and A.~Zisserman,
\newblock ``The conversation: {D}eep audio-visual speech enhancement,''
\newblock in {\em Proc. Conference of the International Speech Communication
  Association (INTERSPEECH)}, 2018, pp. 3244--3248.

\bibitem{GabbSP18}
A.~Gabbay, A.~Shamir, and S.~Peleg,
\newblock ``Visual speech enhancement,''
\newblock in {\em Proc. Conference of the International Speech Communication
  Association (INTERSPEECH)}, 2018, pp. 1170--1174.

\bibitem{sadeghiLAGH19}
M.~Sadeghi, S.~Leglaive, X.~Alameda-Pineda, L.~Girin, and R.~Horaud,
\newblock ``Audio-visual speech enhancement using conditional variational
  auto-encoders,''
\newblock {\em IEEE/ACM Transactions on Audio, Speech and Language Processing},
  vol. 28, pp. 1788--1800, 2020.

\bibitem{SadeA19a}
M.~Sadeghi and X.~Alameda-Pineda,
\newblock ``Robust unsupervised audio-visual speech enhancement using a mixture
  of variational autoencoders,''
\newblock in {\em Proc. IEEE International Conference on Acoustics, Speech, and
  Signal Processing (ICASSP)}, 2020.

\bibitem{SadeA21Switching}
M.~Sadeghi and X.~Alameda-Pineda,
\newblock ``Switching variational auto-encoders for noise-agnostic audio-visual
  speech enhancement,''
\newblock in {\em Proc. IEEE International Conference on Acoustics, Speech, and
  Signal Processing (ICASSP)}, 2021.

\bibitem{bishop06}
C.~Bishop,
\newblock {\em Pattern Recognition and Machine Learning},
\newblock Springer-Verlag Berlin, Heidelberg, 2006.

\bibitem{vincent2006performance}
E.~Vincent, R.~Gribonval, and C.~F{\'e}votte,
\newblock ``Performance measurement in blind audio source separation,''
\newblock {\em IEEE Transactions on Audio, Speech, and Language Processing},
  vol. 14, no. 4, pp. 1462--1469, 2006.

\bibitem{rix2001perceptual}
A.~W. Rix, J.~G. Beerends, M.~P. Hollier, and A.~P. Hekstra,
\newblock ``{Perceptual evaluation of speech quality (PESQ)-a new method for
  speech quality assessment of telephone networks and codecs},''
\newblock in {\em Proc. IEEE International Conference on Acoustics, Speech, and
  Signal Processing (ICASSP)}, 2001, pp. 749--752.

\bibitem{taal2011algorithm}
C.~H. Taal, R.~C. Hendriks, R.~Heusdens, and J.~Jensen,
\newblock ``An algorithm for intelligibility prediction of time--frequency
  weighted noisy speech,''
\newblock {\em IEEE Trans. Audio, Speech, Language Process.}, vol. 19, no. 7,
  pp. 2125--2136, 2011.

\bibitem{Abde17}
A.-H. Abdelaziz,
\newblock ``{NTCD-TIMIT}: A new database and baseline for noise-robust
  audio-visual speech recognition,''
\newblock in {\em Proc. Conference of the International Speech Communication
  Association (INTERSPEECH)}, 2017, pp. 3752--3756.

\bibitem{TIMIT}
M.~S. Garofolo, L.~F. Lamel, W.~M. Fisher, J.~G. Fiscus, D.~S. Pallett, N.~L.
  Dahlgren, and V.~Zue,
\newblock ``{TIMIT} acoustic phonetic continuous speech corpus,''
\newblock in {\em Linguistic data consortium}, 1993.

\bibitem{kingma2014adam}
D.~P. Kingma and J.~Ba,
\newblock ``Adam: A method for stochastic optimization,''
\newblock in {\em International Conference on Learning Representations (ICLR)},
  2015.

\bibitem{girin2020dynamical}
L.~Girin, S.~Leglaive, X.~Bie, J.~Diard, T.~Hueber, and X.~Alameda-Pineda,
\newblock ``Dynamical variational autoencoders: A comprehensive review,''
\newblock {\em arXiv preprint arXiv:2008.12595}, 2020.

\bibitem{FevotBD09}
C.~F{\'e}votte, N.~Bertin, and J.-L. Durrieu,
\newblock ``{Nonnegative matrix factorization with the Itakura-Saito
  divergence: With application to music analysis},''
\newblock {\em Neural computation}, vol. 21, no. 3, pp. 793--830, 2009.

\end{thebibliography}

\begin{IEEEbiography}[{\includegraphics[width=1in,height=1.25in,clip,keepaspectratio]{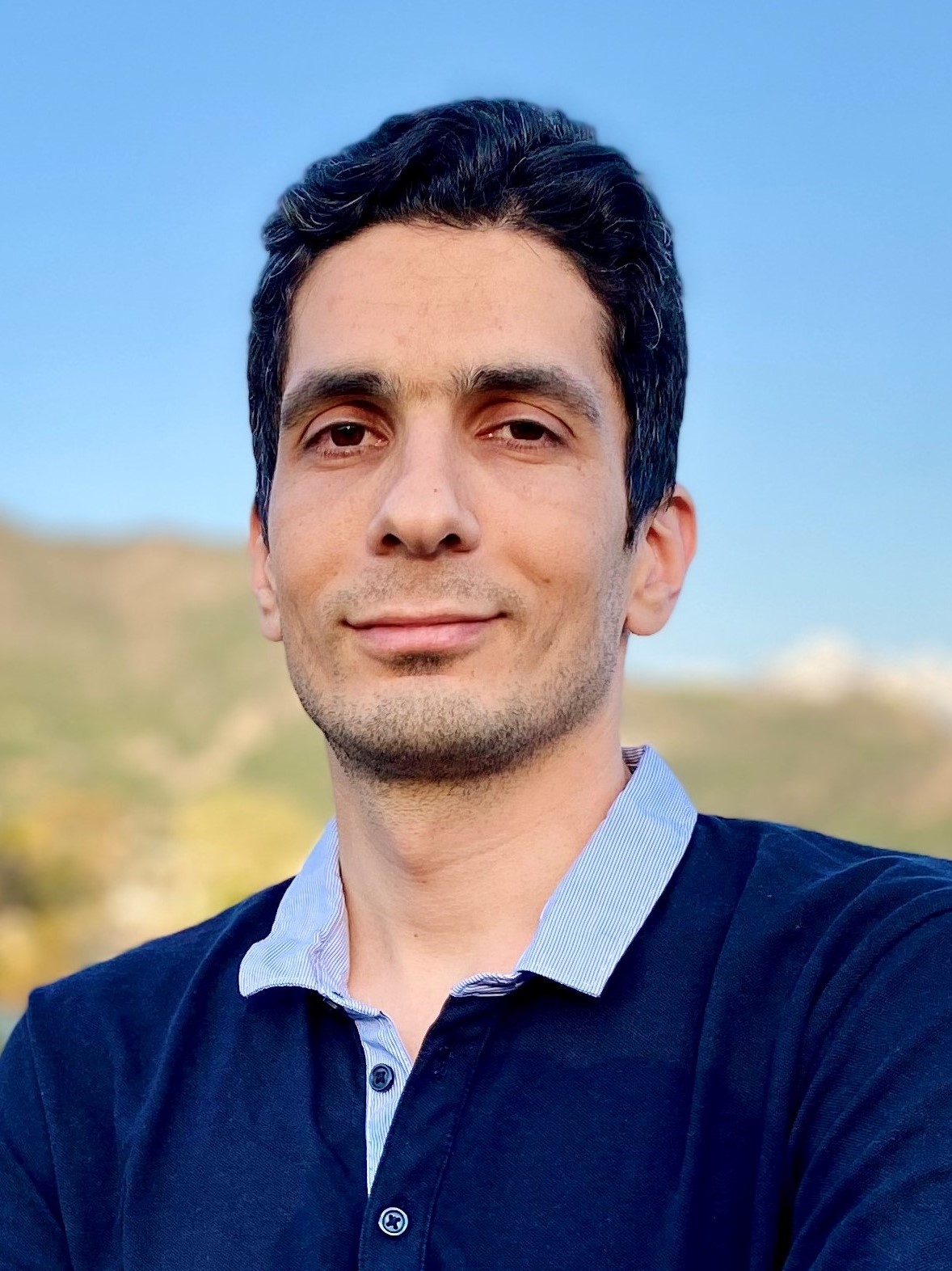}}]{Mostafa Sadeghi} 
	received the B.Sc. degree from Ferdowsi University of Mashhad, Iran, in 2010, the M.Sc. degree from Sharif University of Technology, Tehran, Iran, in 2012, and the Ph.D. degree from the same university in 2018, all in electrical engineering. From August 2018 to October 2020 he was a postdoctoral researcher in the Perception team at Inria Grenoble Rh\^one-Alpes. Currently, he is a researcher in the Multispeech team at Inria Nancy - Grand Est. His main research interests are latent variable generative models, unsupervised audio-visual speech enhancement, Bayesian inference and probabilistic machine learning, and local/global optimization.
\end{IEEEbiography}

\begin{IEEEbiography}[{\includegraphics[width=1in,height=1.25in,clip,keepaspectratio]{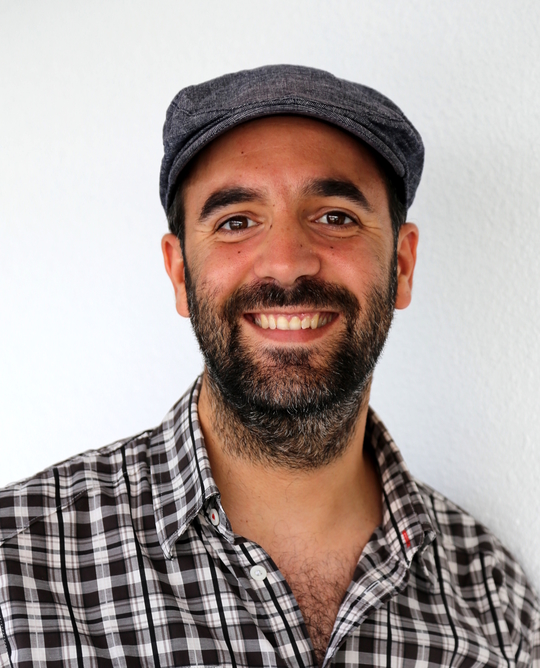}}]{Xavier Alameda-Pineda} is a (tenured) Research Scientist at Inria, in the Perception Group. He obtained the M.Sc. (equivalent) in Mathematics in 2008, in Telecommunications in 2009 from BarcelonaTech and in Computer Science in 2010 from Université Grenoble-Alpes (UGA). He the worked towards his Ph.D. in Mathematics and Computer Science, and obtained it 2013, from UGA. After a two-year post-doc period at the Multimodal Human Understanding Group, at University of Trento, he was appointed with his current position. Xavier is an active member of SIGMM, and a senior member of IEEE and a member of ELLIS. He is co-chairing the “Audio-visual machine perception and interaction for companion robots” chair of the Multidisciplinary Institute of Artificial Intelligence. Xavier is the Coordinator of the H2020 Project SPRING: Socially Pertinent Robots in Gerontological Healthcare. Xavier’s research interests are in combining machine learning, computer vision and audio processing for scene and behavior analysis and human-robot interaction.
\end{IEEEbiography}
	
\end{document}